\documentclass{aa}  

\usepackage{graphicx}
\usepackage{txfonts}
\usepackage{lipsum}
\usepackage{subcaption}
\usepackage{lscape}
\usepackage{placeins} 
\usepackage{bm, amsmath}
\usepackage{orcidlink}

\newcommand{\athena}{\textsf{Athena++}}
\newcommand{\amrvac}{\textsf{MPI-AMRVAC}}
\newcommand{\lare}{\textsf{Lare}}
\newcommand{\pluto}{\textsf{PLUTO}}
\newcommand{\legolas}{\textsf{Legolas}}
\newcommand{\pylbo}{\textsf{pylbo}}
\newcommand{\gimli}{\textsf{GIMLI}}
\newcommand{\yt}{\textsf{yt}}
\newcommand{\pyvista}{\textsf{pyvista}}
\newcommand{\agile}{\textsf{AGILE}}

\newcommand{\bfb}{\bm{B}}

\newcommand{\bfk}{\bm{k}}

\newcommand{\bfg}{\bm{g}}
\newcommand{\ex}{\hat{\bm{e}}_x}
\newcommand{\ey}{\hat{\bm{e}}_y}

\newcommand{\im}{\mathrm{i}}

\begin{document}

\title{\gimli{}: a toolkit connecting linear and non-linear magnetohydrodynamic regimes}

\author{
    J. De Jonghe \inst{1} \corrauth{jordi.dejonghe@kuleuven.be} \orcidlink{0000-0003-2443-3903}
    \and N. Brughmans \inst{1} \email{nicolas.brughmans@kuleuven.be} \orcidlink{0000-0002-7885-4554}
    \and R. Keppens \inst{1} \email{rony.keppens@kuleuven.be} \orcidlink{0000-0003-3544-2733}
    }
\institute{Centre for mathematical Plasma Astrophysics, Department of Mathematics, KU Leuven, Celestijnenlaan 200B, 3001 Leuven, Belgium}

\abstract
% Context (optional)
{}
% Aims (mandatory)
{In this work, we present the General Interface for \amrvac{} / \legolas{} Interconnection (or \gimli{}, for short) as a toolkit for in-depth (magneto)hydrodynamic studies combining linear and non-linear methods.}
% Methods (mandatory)
{Acting as a bridge between the linear (M)HD spectroscopy code \legolas{} and the non-linear simulation code \amrvac{}, \gimli{} facilitates the initialization of non-linear \amrvac{} simulations with linear solutions.}
% Results (mandatory)
{We showcase this functionality on both Cartesian and cylindrical setups, covering representative plasma physical and astrophysical scenarios: Parker unstable magnetized atmospheres, interchange modes in a straight tokamak setup, resistive Harris sheet evolutions and thermally modified discrete Alfv\'en waves in coronal loops. We offer suggestions on how to combine linear and non-linear methods, and demonstrate how to leverage \gimli{} in the process.}
% Conclusions (optional)
{}


\maketitle
\nolinenumbers

\section{Introduction}
Over the past 30 years or so, a large fraction of theoretical research in solar and astrophysics has shifted to the reproduction of observed physical phenomena in fully non-linear, magnetohydrodynamic (MHD) simulations, as evidenced by the (continually ongoing) development of a variety of non-linear MHD simulation codes like \athena{} \citep{Stone2020}, \lare{} \citep{Arber2001}, \amrvac{} \citep{Porth2014, Xia2018, Keppens2023}, and \pluto{} \citep{Mignone2007, Mignone2012}, to name just a few. Through the inclusion (or deliberate exclusion) of various non-ideal terms in the MHD equations, simulations have allowed to study increasingly complicated plasma configurations and how they are affected by the deviations from ideal MHD theory, ultimately leading to a better understanding of phenomena's driving forces.

Despite their remarkable success, their perpetual need for increasingly high resolution to accurately portray smaller-scale structures requires a large amount of computing power. However, sometimes less computationally expensive methods are desirable to gain quicker, and perhaps more fundamental, insight into a configuration's dynamics. In more recent years, this has given rise to a revival of spectroscopic methods to study the linear dynamics of plasma structures, where in this context the term spectroscopy refers to the quantification of a state's natural frequencies and their corresponding perturbations. Notably, \citet{Goedbloed2018a, Goedbloed2018b} achieves this with a novel spectral web technique whereas the \legolas{} code \citep{Claes2020, Claes2023} quantifies a system's  eigenmodes by formulating, and subsequently solving, the generalized eigenvalue problem that arises after writing the linearized MHD equations in a weak Galerkin form. Early examples of this spectroscopic approach for MHD states with flow are found in \citet{Nijboer1997} for 1D cylindrical and plane-parallel setups, or \citet{phoenix2007} for the generalization to 2D axisymmetric rotating plasmas for tokamaks or accretion disks.

Recently, the \legolas{} code was also extended to solve the linear initial-value problem \citep{Kelly2026}, allowing for an accurate representation of the initial, linear stage of evolution. \cite{Kelly2026} shows that for a one-dimensional model of a stratified coronal loop, having exact know-how of the linear eigenmodes and the ensuing linear time evolution can be directly useful to interpret a fully non-linear simulation of the same setup.

So far, both methods have been used mostly in isolation, with non-linear simulations finding applications in a wide variety of astrophysical environments and configurations, and spectroscopic methods traditionally focusing more on fundamental understanding, e.g. waves and instabilities in cylindrical plasmas with resistivity \citep{Kerner1985} or background flow \citep{Nijboer1997}, or, since the conception of the \legolas{} code, in the investigation of thermal instability in the solar atmosphere \citep{Claes2021}, the interaction of magnetic and velocity shear \citep{DeJonghe2024}, and non-axisymmetric instabilities in accretion disks \citep{Brughmans2024,Brughmans2025}. However, the two methods were used alongside each other in \citet{DeJonghe2025}, where a preliminary linear analysis of the tearing and thermal instabilities of a Harris current sheet equilibrium directly informed the initial perturbation used in the non-linear simulation of that same equilibrium, i.e. the initial perturbations of the non-linear simulations were solutions of the linearized system. Following this work, \citet{DeJonghe2026a} demonstrated that this method of initialization provides a tangible numerical advantage by reducing the time spent in the linear stage of the simulation. Here, we present the General Interface for \amrvac{} / \legolas{} Interconnection (abbreviated \gimli{}), a toolkit providing a framework for the initialization of non-linear simulations with user-controlled exact (superpositions of) linear eigenmodes. While we demonstrate this here for the open-source non-linear solver \amrvac{}, compatibility with the GPU-driven \textsf{AGILE} \citep{Porth2026} is achieved with minimal adjustments and will be exploited in the near future.

To minimize room for user error, \gimli{} also acts as an interface from which user files for \legolas{} and \amrvac{} can be generated from a single configuration definition.
Secondly, \gimli{} supports going the other way as well, starting from a non-linear simulation snapshot and preparing one-dimensional cuts for analysis with \legolas{}. Though such an analysis only yields sensible results if the gradients perpendicular to the cut are much smaller than along the cut, this method has been successfully applied in a solar streamer setting, where streamer waves were identified in the resulting \legolas{} spectrum \citep{DeJonghe2026b}. That work analyzed spectra from a snapshot prior to a developing streamer wave pattern, but we expect that spectrally analyzing subsequent time-evolving cuts of fully non-linear MHD evolutions may reveal distinct changes in their corresponding eigenfrequency spectrum, an aspect still to be explored fully. This expectation rests on findings that MHD spectral theory may well dictate the entire non-linear evolution, as noted in a field theoretical setting of ideal MHD by \citet{fieldtheory2016}.

In this work, we demonstrate \gimli{}'s simulation initialization capabilities and offer suggestions on possible applications. Sec. \ref{sec:overview} offers a short overview of the \legolas{} and \amrvac{} codes, and how \gimli{} interacts with both. Then, Sec. \ref{sec:examples} demonstrates how to set up problems simultaneously in \legolas{} and \amrvac{} with \gimli{} and how to initialize simulations with linear solutions. Finally, we provide our perspective on the advantages of combining linear and non-linear tools in Sec. \ref{sec:outlook}.

\section{Overview}\label{sec:overview}
Before presenting the novel \gimli{} toolkit, we briefly summarize what the \legolas{} and \amrvac{} codes do. Then, a summary of \gimli{}'s purpose and functionality is presented. Finally, we describe a typical \gimli{} workflow, as used in the examples of Sec. \ref{sec:examples}.

\subsection{The \legolas{} and \amrvac{} codes}
\legolas{} is an MHD spectroscopic code that calculates a configuration's natural oscillations and instabilities for a given Cartesian or cylindrical (equilibrium) state. Concretely, the code assumes a one-dimensionally varying state (depending either on $x$ or $r$) around which the MHD equations are perturbed and linearized, modularly including non-ideal terms if desired \citep[for an overview of all terms, see \url{https://legolas.science};][]{Claes2020, DeJonghe2022}. Then, the perturbed quantities $f_1$ (density $\rho$, velocity $\bm{v}$, temperature $T$, and magnetic field $\bm{B}$) are all assumed to follow an exponential time ($t$) dependence $f_1 \sim\exp(-\mathrm{i}\omega t)$ and Fourier analysis is applied in the two homogeneous directions, i.e. $f_1 \sim \exp(\mathrm{i} k_2 u_2 + \mathrm{i} k_3 u_3)$. As is common, $\omega$ and $k_j$ ($j = 2,3$) represent angular frequency and wavenumbers, respectively. Here, $u_2$ and $u_3$ correspond to the $y$- and $z$-coordinates in Cartesian configurations, or $\theta$ and $z$ in a cylindrical setup. By writing the resulting equations in a weak Galerkin form the problem is reduced to a generalized eigenvalue problem where the eigenvalues represent frequencies $\omega$ and the eigenvectors correspond to the frequencies' associated perturbation amplitudes, which are functions of $x$ or $r$. Solving the eigenproblem we thus obtain a spectrum of frequencies whose real parts describe oscillatory behavior, imaginary parts the growth/damping rate, and associated eigenfunctions $x$- or $r$-dependent perturbation amplitudes. We note that both eigenfrequencies and their associated eigenfunctions are defined as complex-valued functions, as we switched to the familiar Fourier representations.

\amrvac{} on the other hand is a parallelized generic partial differential equation solver, including adaptive mesh refinement capabilities, for a variety of systems of (primarily hyperbolic) partial differential equations (PDE), but in particular the hydrodynamic and magnetohydrodynamic models. With a choice from its assortment of time integration procedures, the code calculates a discretized temporal evolution of a user-defined initial configuration in 1 to 3 dimensions. Data is saved at regular intervals, defining the density, pressure, velocity, and magnetic field at a given time. With some post-processing the dynamics of the configuration can be visualized and analyzed as a function of time, and even compared to observations. As usual for generic non-linear MHD solvers, this code operates in real space only, and advances the PDE system forward in time.

\subsection{\gimli{} summary}
Traditionally, numerically reproducing time evolutions of linearly unstable equilibria is ensured by adding a small perturbation to the equilibrium, either following some analytical prescription or by introducing random noise. However, as shown in \citet{DeJonghe2026a}, there may be a substantial numerical benefit to perturbing with unstable linear solutions as it allows the evolution to skip the typical startup errors and initial transients that do not correspond to eigenmodes. Secondly, isolating modes allows us to hone in on their specific influence on the system's dynamics, but also better understand the coupling that occurs between different modes. E.g. in the gyrokinetics community, there is a lot of attention to the importance of stable modes in driving turbulence and how a strongly turbulent state retains information on the underlying linear eigenmodes \citep{Hatch2016, Pueschel2016, Fraser2018}.

To facilitate the use of linear solutions as initial perturbations in non-linear simulations, the open-source \gimli{} toolkit provides a framework to set up identical configurations in \legolas{} and \amrvac{}, and insert linear solutions from \legolas{} in \amrvac{} simulations. Written in Python, \gimli{} is distributed as part of the \pylbo{} Python package, available from the \legolas{} repository. Relying on \textsf{SymPy} \citep{Meurer2017}, the initial equilibrium is defined as a set of analytical, symbolic expressions. In \gimli{}, these expressions are converted to Fortran code to generate user modules for both \legolas{} and \amrvac{}. In addition, after identifying the linear modes of interest in \legolas{} data, \gimli{} calculates the desired weighted superposition and writes it to a file that can be read by \amrvac{}.

\subsection{Workflow}\label{sec:workflow}
\gimli{} is designed to facilitate a workflow in which linear results directly inform non-linear simulations, from start to finish. First, the one-dimensionally varying equilibrium of interest is defined as a set of \textsf{SymPy} expressions in \gimli{}. Then, a dictionary is constructed for \legolas{} specifying the values of any parameters in the equilibrium, the geometry and domain, the resolution, which physics to include, and any other \legolas{} parameters you wish to set manually (e.g. which solver to use or output to save). The dictionary borrows its structure from \legolas{}'s multirun framework in order to facilitate parameter scans, and in particular, variation in wavenumber. With both the equilibrium and configuration dictionary, \gimli{} will generate a \legolas{} user module and parameter file(s), for use with \legolas{} v2.3.1 or later.

After compiling \legolas{} and running the code with the generated parfile(s), the spectrum is visualized with \pylbo{} and the relevant modes are identified. The \gimli{} equilibrium object and the selected eigenvalues are then added to a dictionary to inform the \amrvac{} setup. Further, the geometry and equilibrium parameters are copied from \legolas{}'s dictionary, and the simulation specifications are detailed, defining the dimensionality and the simulation bounds in each direction. Here, the bounds for the first coordinate should match those used in \legolas{}, whereas in the examples below the bounds in the other directions are informed by the wavenumbers of the \legolas{} runs, such that the simulation box size accommodates integer multiples of the inserted wavelengths. Naturally, the dictionary should further specify how the amplitude of the perturbation is set compared to the equilibrium, either based on a background field's maximum value or the increase in total energy. Finally, the dictionary should contain all the simulation specifics, such as resolution, integration method, and output cadence. From this information, \gimli{} is applied to generate a data file containing the perturbation arrays, and an \amrvac{} parameter file and user module compatible with \amrvac{} v3.3 or later. The latter features the equilibrium profiles and the necessary routines to read the newly-generated perturbation file, and interpolate the arrays to \amrvac{}'s grid. Optionally, the user can enable splitting off the time-invariant equilibria for the magnetic field and current density, as well as density and pressure, so that only the perturbed variables are time-integrated. Alternatively, routines are available to apply boundary conditions only to the perturbation by splitting off the equilibria. After compiling \amrvac{} with this user module, the simulation is run and analyzed as usual, e.g. with the Python package \yt{} \citep{Turk2011}, as done in the next section.

This workflow is demonstrated for four examples in Sec. \ref{sec:examples}. The process is further documented with code snippets on the website (\url{https://legolas.science}), and the complete workflows for the Harris current sheet and discrete Alfvén setup are available in the code's GitHub repository as an example. Note that everything is presented in dimensionless form, because the results here are independent of the normalization, unless indicated otherwise.

\section{Demonstration}\label{sec:examples}
In this section we present four examples of the \gimli{} workflow described in Sec. \ref{sec:workflow}. To cover a variety of dynamics, we consider two Cartesian and two cylindrical configurations, one of each in ideal MHD and one of each including non-ideal effects. The non-ideal effects considered in these examples are resistivity, optically thin radiative cooling and parallel thermal conduction. Additionally, different dimensionalities are selected for the corresponding \amrvac{} simulations. Though \gimli{} is fully compatible with \amrvac{}'s adaptive mesh refinement (AMR) capabilities, resolutions were kept fixed for this demonstration. 

\subsection{Ideal MHD}
Let us first consider ideal MHD. As noted, we present two examples, one Cartesian and one cylindrical. In Cartesian geometry we study an unstable, gravitationally stratified atmosphere and in cylindrical geometry an approximation of a constant-current tokamak.

\subsubsection{Parker instability}
\begin{figure}
    \centering
    \includegraphics[width=\linewidth]{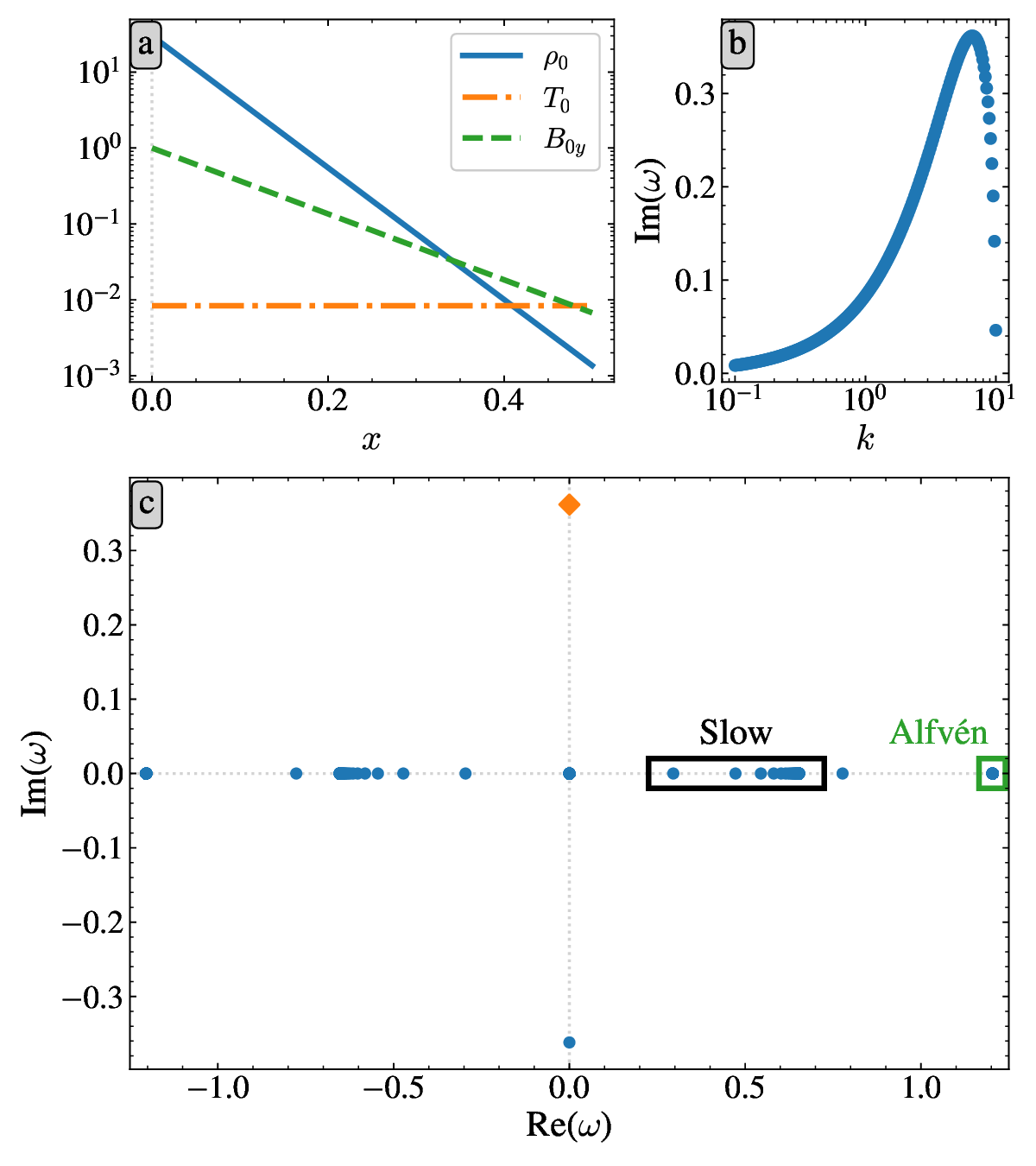}
    \caption{(a) Gravitationally stratified atmosphere equilibrium profiles. (b) Parker growth rate as a function of $k$. (c) Spectrum of eigenfrequencies for $k_\mathrm{max} \simeq 6.59$. The Parker instability is represented by an orange diamond.}
    \label{fig:legolas-parker}
\end{figure}

As a first example, we consider a static, gravitationally stratified atmosphere with a horizontal magnetic field $\bfb_0$. Assuming a constant gravitational acceleration $\bfg = -g\,\ex$, the equilibrium takes the form
\begin{equation}
\begin{aligned}
    &\rho_0(x) = \rho_c \exp(-\alpha x), &&p_0(x) = p_c \exp(-\alpha x), \\ &\bfb_0(x) = B_c \exp\left(-\frac{\alpha x}{2}\right)\,\ey, \quad &&\alpha = \frac{\rho_c g}{p_c + B_c^2/2},
\end{aligned}
\end{equation}
where this combination of density $\rho_0$ and thermal pressure $p_0$ yields a uniform temperature $T_0 = p_0/\rho_0 = p_c / \rho_c$, and the expression for $\alpha$ is obtained from the force-balance equation
\begin{equation}
    \left( p_0 + \frac{1}{2} \bfb_0^2 \right)' + \rho_0 g = 0.
\end{equation}
This configuration is unstable to the Parker instability \citep[see e.g.][]{Goedbloed2019} with a wave vector $\bfk = k\,\ey$ parallel to the magnetic field. Here, we consider the interval $x\in [0, 0.5]$ and the parameters are set to $g = 0.5$, $\alpha = 20$, $B_c = 1$, and $p_c = 0.25$, yielding $\rho_c = 30$ and $T_0 \simeq 8.33 \times 10^{-3}$. For these values, the equilibrium profiles are shown on a logarithmic scale in Fig. \ref{fig:legolas-parker}a.

With a uniform grid of $351$ grid points, the wavenumber of maximal growth rate is approximated by performing $200$ \legolas{} runs with $k$ ranging from $0.1$ to $10$, using inverse vector iteration to find the instability. The growth rate as a function of $k$ is shown in Fig. \ref{fig:legolas-parker}b, with the maximum in the series occurring at $k_\mathrm{max} \simeq 6.59$. Although the wavenumber of maximal growth could be further refined with a finer discretization in $k$, this suffices for our demonstration. The central part of this wavenumber's full spectrum, calculated with a QR algorithm using a Cholesky decomposition, can be seen in Fig. \ref{fig:legolas-parker}c, with the Parker instability indicated with an orange diamond. In addition, the sequence of forward propagating slow modes has been indicated with a black rectangle, as well as the Alfv\'en waves in green, noting that the Alfv\'en speed is constant throughout the slab.

\begin{figure}
    \centering
    \includegraphics[width=\linewidth]{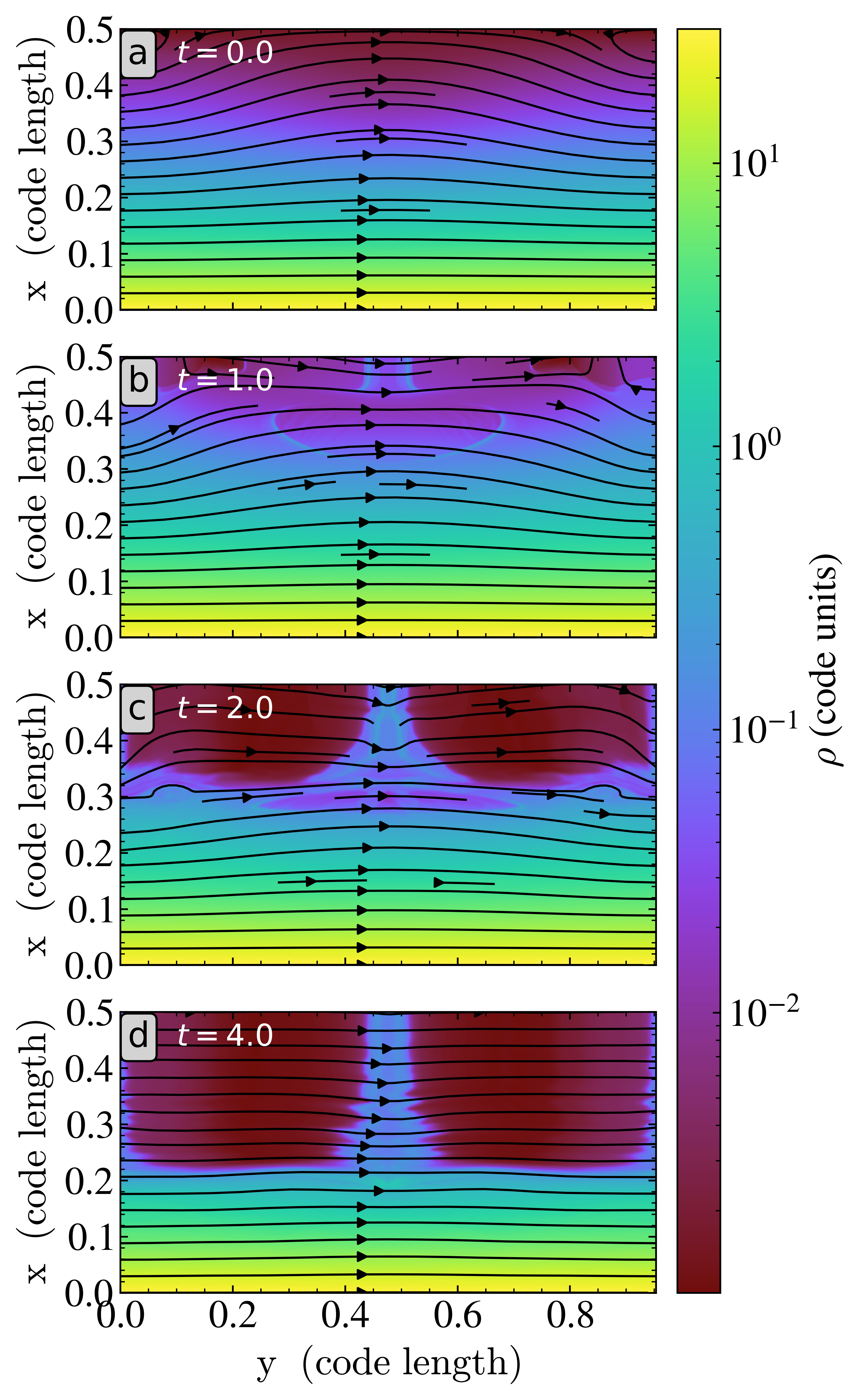}
    \caption{Density and magnetic field lines of the gravitationally stratified atmosphere at (a) $t = 0$, (b) $t = 1$, (c) $t = 2$, and (d) $t = 4$.}
    \label{fig:amrvac-parker}
\end{figure}

Setting up a 2D \amrvac{} simulation with \gimli{}, we match the domain size in the $x$-direction to the \legolas{} configuration, i.e. $x\in[0,0.5]$, and in the $y$-direction we take $y\in[0, \lambda_\mathrm{max}]$ for $\lambda_\mathrm{max} = 2\pi/k_\mathrm{max} \simeq 0.953$, discretized with $512\times 1024$ cells. Like in \legolas{}, we apply a perfectly conducting wall boundary condition at $x=0$, but we change the boundary condition at $x=0.5$ to a continuous one for all quantities. Most notably, this will allow the magnetic field to develop a component perpendicular to the top boundary after the initial time. Periodic boundary conditions are applied in the $y$-direction. Then, the fastest growing mode is added to the equilibrium with a maximal amplitude of $0.01\,B_c$ in the $B_y$-perturbation. The resulting configuration is shown in Fig. \ref{fig:amrvac-parker}a, colored by density and annotated with the magnetic field lines. The other three panels show the situation at later times, evolved using a three-step time integration with an HLL flux scheme and WENO5 limiter, where the atmosphere has moved further away from a simple stratification, until the bent field lines are straightened out again by a density stratification in the $y$-direction.

\begin{figure}
    \centering
    \includegraphics[width=\linewidth]{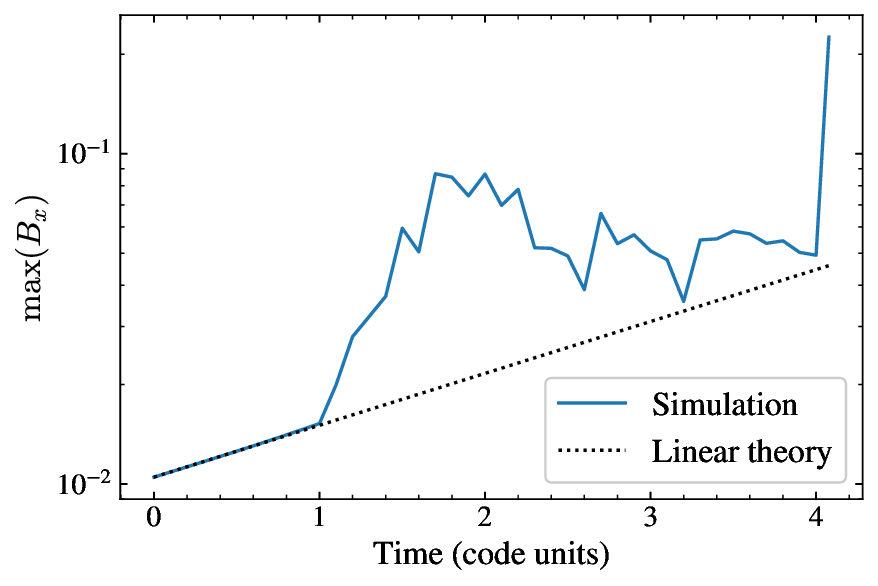}
    \caption{Evolution of $\max(B_x)$ in the simulation (blue) compared to the linear growth rate of the initializing Parker instability (black dashed).}
    \label{fig:linear-parker}
\end{figure}

Since we have both a linear prediction for the initial growth rate and a non-linear simulation, we can pinpoint when the simulation deviates from linear behavior. Since the Parker instability introduces a non-zero $B_x$-component, which the equilibrium lacks, $\max(B_x)$ provides a good measure to investigate the growth rate. In Fig. \ref{fig:linear-parker}, the maximal $B_x$-value in the simulation is shown as a function of time alongside the linear prediction $\max(B_x(t=0)) \exp(-\im\omega_\mathrm{max}t)$. It is observed that the simulation starts to deviate significantly from the linear prediction after $t\sim 1$, where it features a sharp jump. At this point, the continuous top boundary condition is clearly observed to influence the further evolution, as evidenced by the shocks in Fig. \ref{fig:amrvac-parker}b that originate at the center of the top boundary. The magnetic field at the top has also clearly developed a component perpendicular to the boundary away from the boundary center. Subsequently, the atmosphere is seen to collapse to form a central column of higher density flanked by low-density regions (Fig. \ref{fig:amrvac-parker}c), causing the magnetic field lines to straighten out again (Fig. \ref{fig:amrvac-parker}d).

\subsubsection{Tokamak constant current} \label{sec:tokamak}
\begin{figure}
    \centering
    \includegraphics[width=\linewidth]{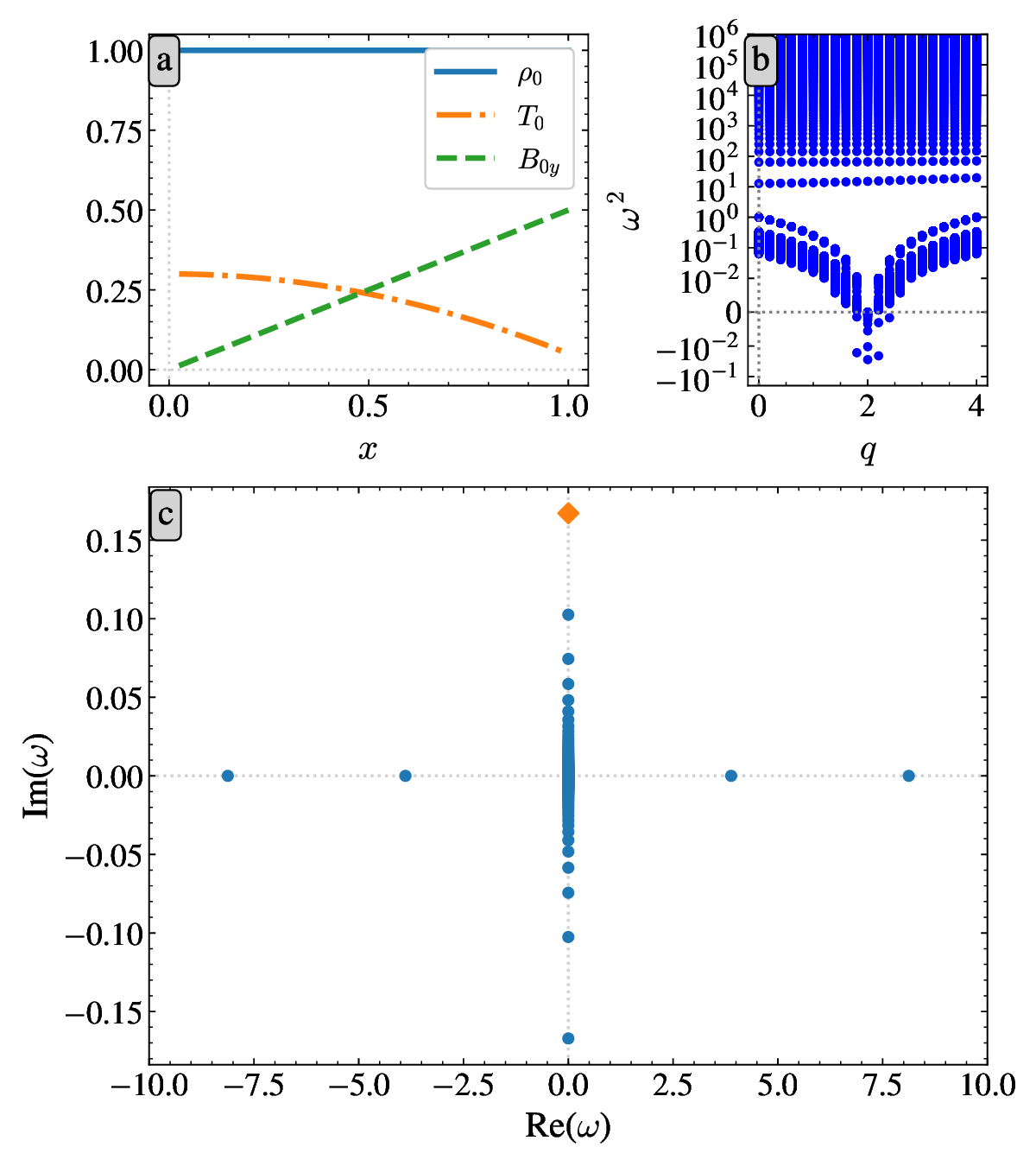}
    \caption{(a) Tokamak constant current equilibrium profiles. (b) $\omega^2$ as a function of safety factor $q=2n/(j_0 R)$, with $R=10$ and $j_0=1$. (c) Spectrum of eigenfrequencies for $k_\mathrm{max} = 1$. The most unstable interchange mode is represented by an orange diamond.}
    \label{fig:legolas-tokamak}
\end{figure}

\legolas{} and \amrvac{} also support cylindrical models that represent e.g. a straight approximation of coronal loops and tokamaks, or an accretion disk or astrophysical jet. As a second test case, we use a constant-current tokamak equilibrium, which is susceptible to various ideal MHD instabilities \citep{Kerner1985,Claes2020}. The radially-varying equilibrium is fixed by assuming $\rho_0 = 1$, $B_{z0} = 1$, and a constant current $j_0$ in the vertical direction. The latter condition implies the following radial profile for the azimuthal field, whereas radial force balance, now including the magnetic curvature $B_\theta/r$, fixes the pressure profile up to a constant at the outer edge $r=1$:
\begin{equation} \label{tokamak:equilibrium}
    B_{\theta0}(r) = \frac{1}{2}j_0 r, \quad p(r) = \frac{1}{4} j_0^2 (1 - r^2) + 0.05.
\end{equation}
As in \cite{Kerner1985}, we assume that the domain represents a section of a tokamak with major radius $R=10$ and minor radius $a=1$. Our vertical wavenumber $k$ in \legolas{} is then connected to the toroidal wavenumber $n$ in a tokamak through the wavelength $\lambda$:
\begin{equation} \label{tokamak:wavenumbers}
    \frac{2\pi R}{n} = \lambda = \frac{2\pi}{k} \quad\Leftrightarrow\quad k = \frac{n}{R}.
\end{equation}
A well-known result \citep{Goedbloed2019} is that the most unstable mode with azimuthal wavenumber $m$ obeys the following relation involving the safety factor $q$:
\begin{equation} \label{tokamak:safety_factor}
    q(r) = \frac{r B_z}{R B_\theta} = \frac{2k}{j_0} = -m.
\end{equation}
We fix $m=-2$ and $j_0=1$, which implies that the most unstable mode has $k=1$ (and hence $n=10$). This choice ensures that $\mathbf{k}\cdot\mathbf{B}=k B_z+m B_\theta/r=0$ throughout the tokamak and the unstable modes are interchange instabilities. This is confirmed by a \legolas{} parameter scan in Fig.~\ref{fig:legolas-tokamak} where we vary $n$ in discrete values, performed with 200 grid points per run on a domain $r\in[0.025, 1]$. The most unstable mode has a growth rate of $\omega_i \approx 0.167$.

We set up a 3D cylindrical simulation in \amrvac{} using \gimli{} taking the eigenfunctions for $k=1$ as initial conditions in the \amrvac{} simulation, where we extend the domain to include the pole by extrapolating the eigenfunctions for regularity (related to their Frobenius expansion near $r=0$ \citep{Goedbloed2019}):
\begin{equation} \label{tokamak:frobenius}
\begin{cases}
    f \sim r^{|m|}, & |m| \neq 0, \\
    f \sim r^2 & m=0.
\end{cases}
\end{equation}
The domain then has $r\in[0,1]$, $\theta\in[0, 2\pi]$, and $z\in[0, 2\pi/k]$, with a resolution of $256\times256\times64$. The lower resolution in the vertical direction does not impact the initial growth significantly since the mode is not propagating. The boundary condition is set to be $\pi$-periodic at the pole, and periodic in both the azimuthal and vertical direction. At the outer radial boundary, we apply the perfectly conducting wall boundary condition from \legolas{} and extrapolate the parallel magnetic field. We use a three-step time integration with the HLL flux solver and cada3 limiter, solving only for the perturbed variables by splitting off a time-invariant background for $\mathbf{B_0}$, $\mathbf{j_0}$, $p_0$, and $\rho_0$.
\begin{figure}
    \centering
    \includegraphics[width=\linewidth]{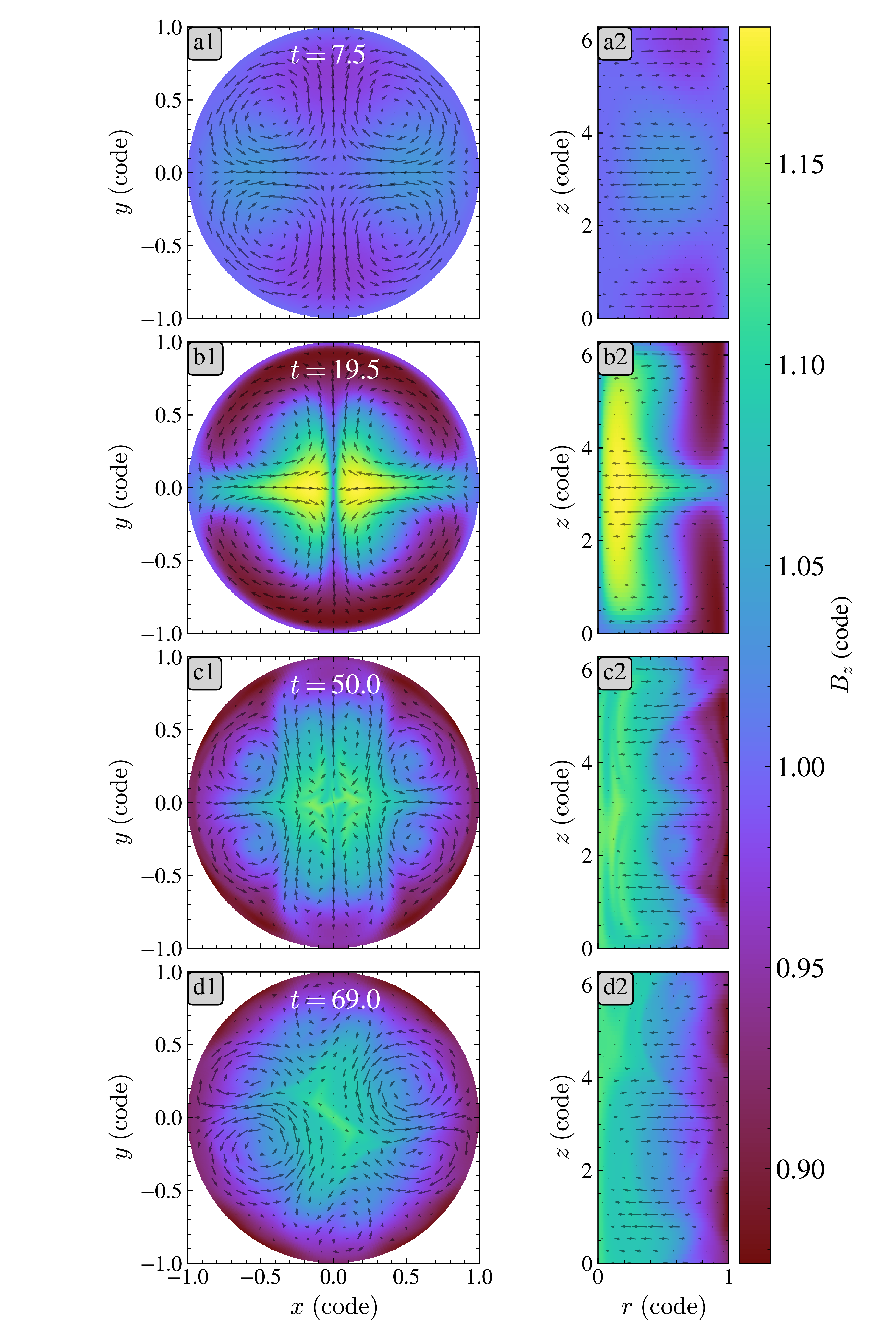}
    \caption{$B_z$ and in-plane velocity field at $z=\pi$ (horizontal) and $\theta=\pi$ (vertical) for (a) $t = 7.5$, (b) $t = 19.5$, (c) $t = 50$, and (d) $t = 69$. An animated version of this figure is available in the online manuscript.}
    \label{fig:amrvac-tokamak}
\end{figure}

\begin{figure}
    \centering
    \includegraphics[width=\linewidth]{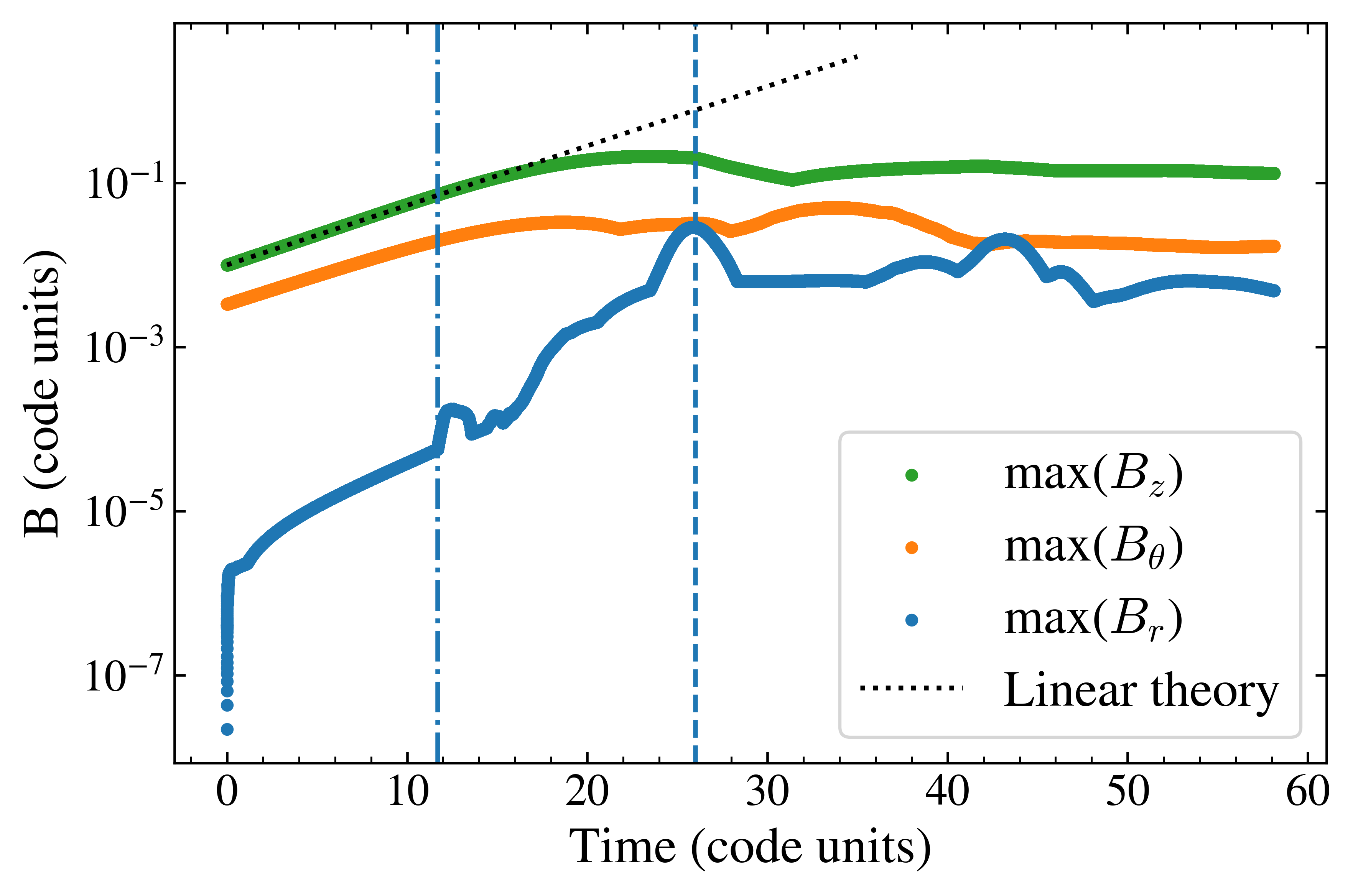}
    \caption{Evolution of the maximum perturbed magnetic fields in the simulation compared to the linear growth rate of the initializing interchange (dashed line). The evolution of $B_r$ determines where non-linear effects take over and the instability saturates (blue vertical lines).}
    \label{fig:linear-tokamak}
\end{figure}

\begin{figure}
    \centering
    \includegraphics[width=\linewidth]{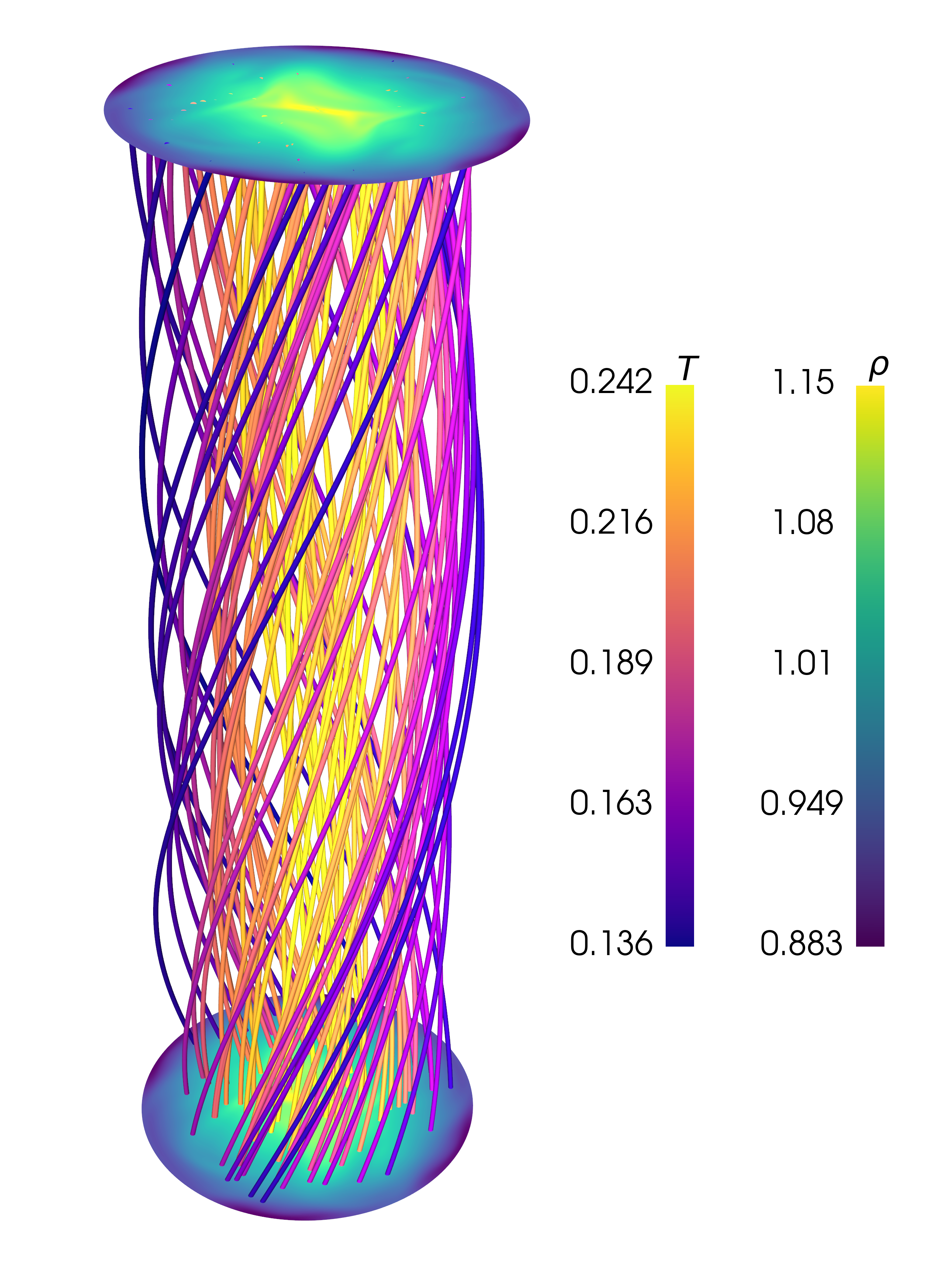}
    \caption{3D \pyvista{} rendering of magnetic field lines at $t=69$ for the interchange instability, colored according to temperature with cuts showing density.}
    \label{fig:tokamak-3D_plot}
\end{figure}

The initial condition perturbs $B_z$ by $1\%$ and is visually almost identical to the first row of Fig.~\ref{fig:amrvac-tokamak}. Other panels show the later evolution and eventual saturation of the instability. Since the mode is purely exponentially growing at a rate of $\omega_i \approx 0.167$, the linear phase of the evolution shows the instability growing in place. Figure~\ref{fig:linear-tokamak} shows the time evolution of the maximum of the perturbed magnetic field, 
with a perfect match with the predicted growth rate. The evolution starts deviating from linear growth around $t=15.5$, when $B_r$ grows quickly. The instability saturates when $B_r$ becomes on par with the value of $B_\theta$ at around $t=26$.  When the perturbation reaches about $10\%$ of the background, smaller-scale structures appear that mixes the material in the tokamak with eruptions emerging from the center of the tokamak (see the animations online). In this case, we find that the initial condition truly excites the most unstable mode, with no additional mode coupling in the linear phase: a Fourier analysis shows that the dominant wavenumbers remain $m=-2$, $k=1$ throughout the linear phase. This indicates that the choice of the initial condition has a large influence on the eventual state here. Figure~\ref{fig:tokamak-3D_plot} shows a 3D visualization of the magnetic field lines at a later time $t=69$, colored according to temperature with cuts showing the density. From the helical field lines it is clear that the equilibrium configuration still dominates after saturation of the instability.

\subsection{Non-ideal MHD}
Now, we consider two non-ideal cases, again one Cartesian and one cylindrical. The Cartesian configuration is a tearing-unstable Harris sheet in resistive MHD. For the cylindrical case, we consider optically thin radiative losses, heating, and parallel thermal conduction in combination with discrete Alfv\'en waves.

\subsubsection{Harris sheet}
\begin{figure}
    \centering
    \includegraphics[width=\linewidth]{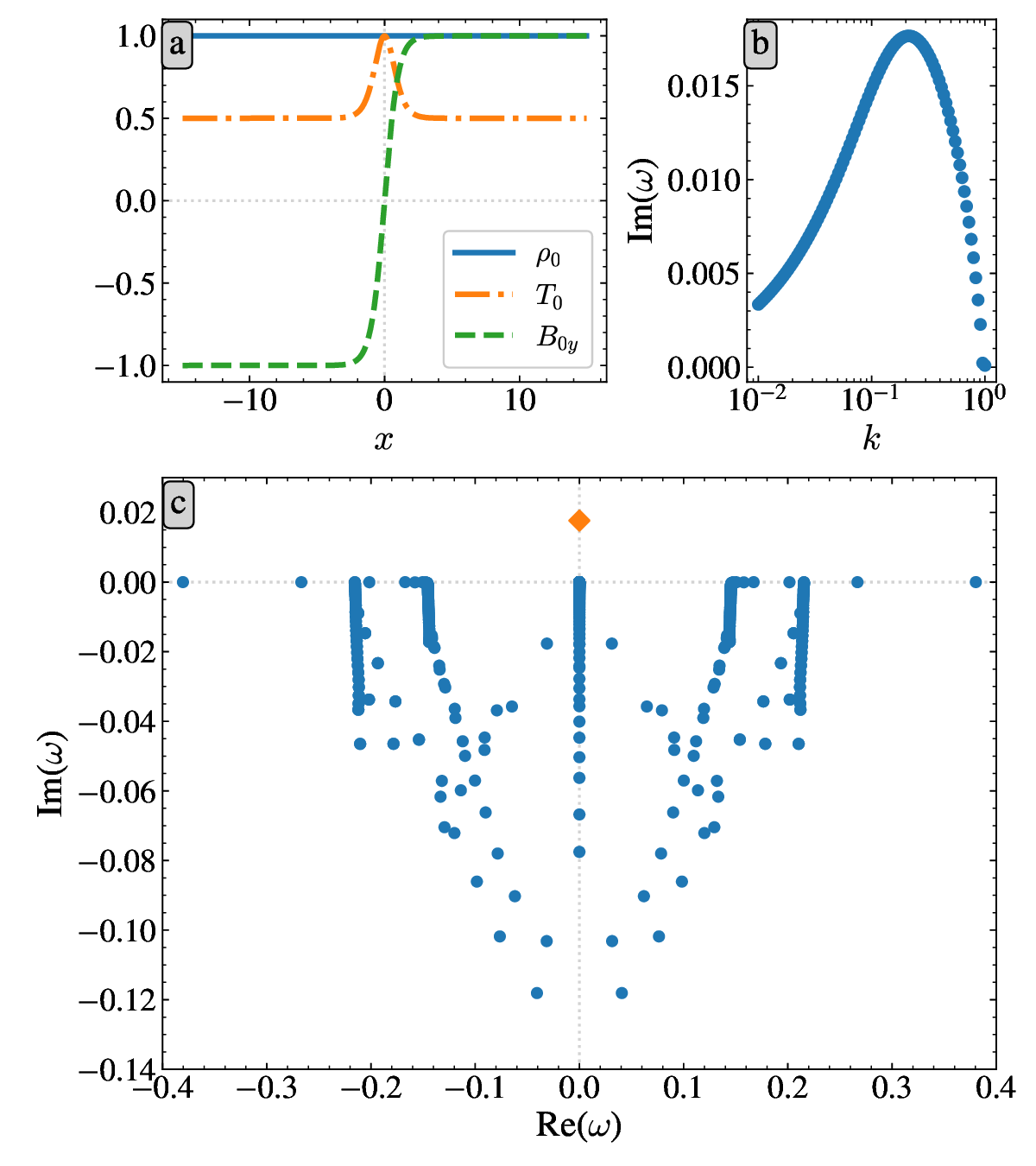}
    \caption{(a) Harris sheet equilibrium profiles. (b) Tearing growth rate as a function of $k$. (c) Spectrum of eigenfrequencies for $k_\mathrm{max} \simeq 0.215$. The tearing instability is represented by an orange diamond.}
    \label{fig:legolas-harris}
\end{figure}

For our example of resistive MHD, we consider a static Harris current sheet, with magnetic field profile
\begin{equation}
    \bfb_0(x) = B_c \tanh\left( \frac{x}{a} \right)\,\ey
\end{equation}
and uniform density $\rho_0$. The force-balance equation is satisfied by adopting the temperature
\begin{equation}
    T_0(x) = T_c - \frac{\bfb_0^2(x)}{2\rho_0}.
\end{equation}
For the purpose of this demonstration, we set all dimensionless parameter values to $1$, i.e. $\rho_0 = T_c = B_c = a = 1$, and add a resistivity of $10^{-3}$. With a non-zero resistivity the Harris sheet is technically not in equilibrium, but the diffusive timescale is much larger than the tearing timescale such that we can treat it as an equilibrium. The resulting equilibrium profiles are shown on the interval $x \in [-15, 15]$ in Fig. \ref{fig:legolas-harris}a.

To discretize this interval in \legolas{}, we use a Gaussian profile
\begin{equation}
    f(x) = p_1 - (p_1-p_3) \exp\left( -\frac{(x-p_2)^2}{2p_4} \right),
\end{equation}
with parameters $p_1 = 0.75$, $p_2=0$, $p_3 = 10^{-3}$, and $p_4=2.5$, to determine the grid spacing as described in \citet{DeJonghe2024}, resulting in $283$ grid points. A perfectly conducting wall boundary condition is applied on either side. Considering this configuration is essentially two-dimensional, we limit the wave vector to the plane, i.e. $\bfk = k\,\ey$, and use the multirun framework to perform a parameter sweep over wavenumber $k$ with $100$ runs utilizing inverse vector iteration to find the tearing mode. The tearing growth rate as a function of $k$ is displayed in Fig. \ref{fig:legolas-harris}b. In this sampling, the maximal growth rate was obtained for $k_\mathrm{max} \simeq 0.215$. Though the wavenumber of maximal growth can be refined further, this suffices for our demonstration. Rerunning the code for this wavenumber with a QR algorithm using Cholesky decomposition yields the full spectrum of eigenvalues, the central part of which is shown in Fig. \ref{fig:legolas-harris}c, with the tearing mode marked by an orange diamond. The inner arc of the damped modes are slow modes whereas the outer arc are Alfv\'en waves. On either side of the spectrum, the first modes of the forward- and backward-propagating fast sequences can be seen, respectively.

\begin{figure}
    \centering
    \includegraphics[width=\linewidth]{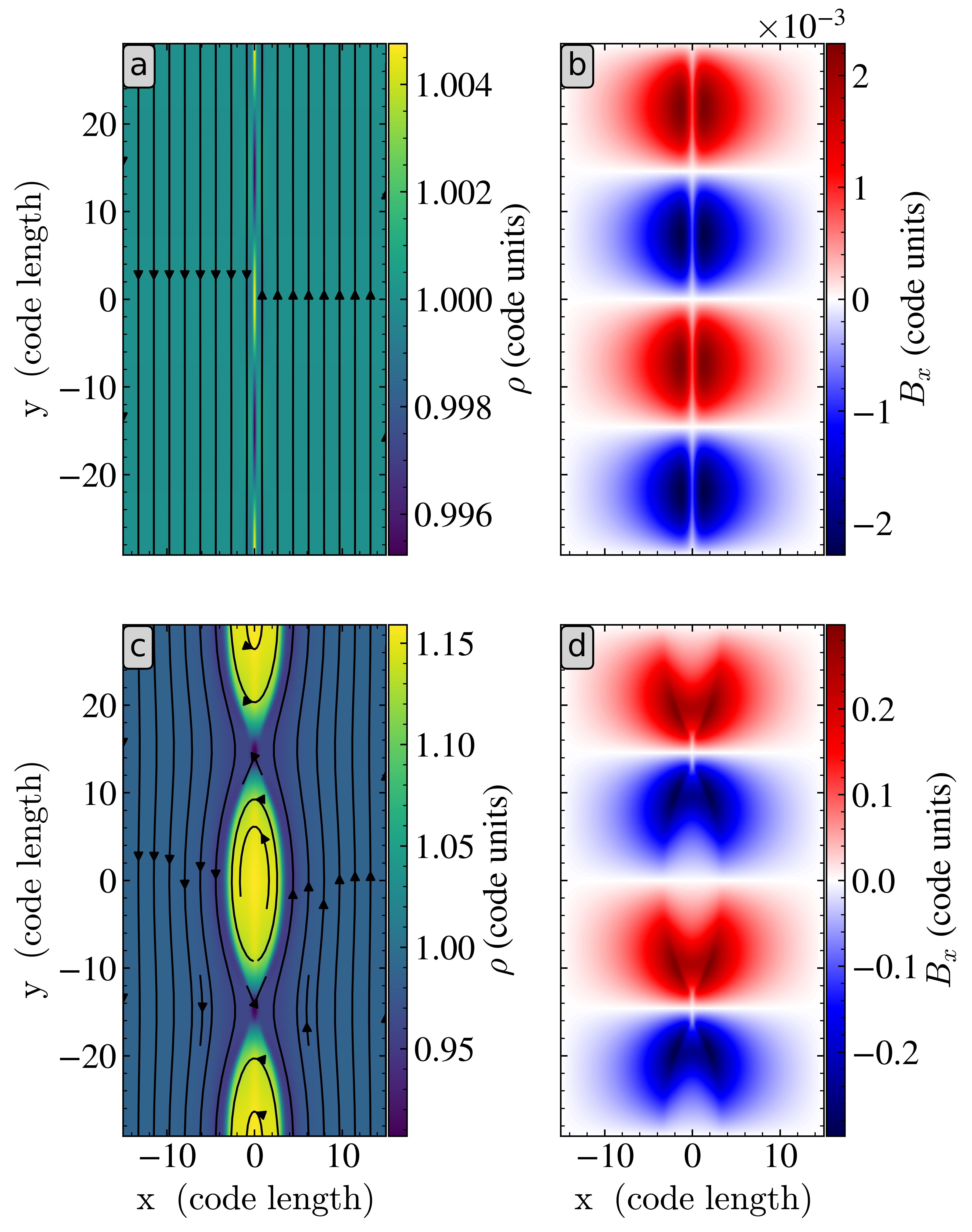}
    \caption{(a) Density at $t = 0$ with annotated magnetic field lines. (b) Magnetic field $B_x$-component at $t = 0$. (c) Density at $t = 100$ with annotated magnetic field lines. (d) Magnetic field $B_x$-component at $t = 300$.}
    \label{fig:amrvac-harris}
\end{figure}

Now, we use \gimli{} to set up a 2D \amrvac{} simulation of that same Harris sheet. In the $x$-direction, the domain size is unaltered from the \legolas{} setup, i.e. $x \in [-15, 15]$, and perfectly conducting wall boundary conditions are applied. In the $y$-direction, the domain is taken to be $y \in [-\lambda_\mathrm{max}, \lambda_\mathrm{max}]$, with $\lambda_\mathrm{max} = 2\pi/k_\mathrm{max}$ the wavelength of the fastest growing mode identified in the \legolas{} study. Periodic boundary conditions are used in this direction. The resulting domain is discretized with $512\times 1024$ cells. Then, the perturbation of the fastest growing tearing mode, calculated with \legolas{}, is added to each respective field. Here, the scaling was determined by demanding that the maximal perturbation in the $B_y$-component is $1\%$ of $B_c$. The initial code density and $B_x$-component after this process are visualized in Figs. \ref{fig:amrvac-harris}a and \ref{fig:amrvac-harris}b, respectively, and the magnetic field lines are annotated in the density panel. The density displays a clear seeding of a density oscillation at the magnetic nullplane, which will grow to form plasmoids. Simultaneously, the $B_x$-component has the typical double-lobed structure of a tearing mode, as expected. Evolving the simulation until $t = 100$ using a three-step time integration with a Total Variation Diminishing Lax-Friedrichs (TVDLF) flux scheme and `minmod' limiter, the density and magnetic field are shown again in Figs. \ref{fig:amrvac-harris}c and \ref{fig:amrvac-harris}d. The $B_x$-component has increased by two orders of magnitude, which is reflected in the well-developed magnetic islands visible in the density panel.

\begin{figure}
    \centering
    \includegraphics[width=\linewidth]{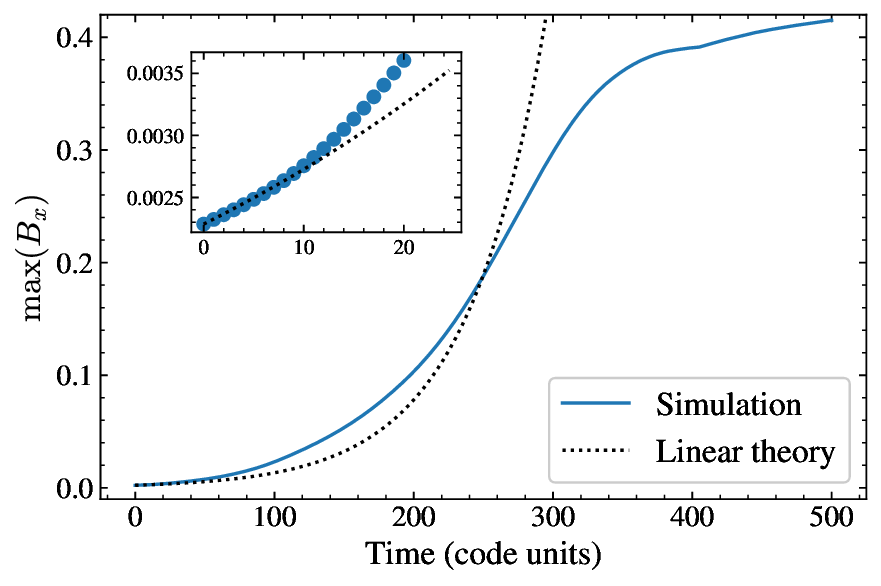}
    \caption{Evolution of $\max(B_x)$ in the simulation (blue) compared to the linear growth rate of the initializing tearing mode (black dashed). The inset shows the initial phase and deviation from linear behavior.}
    \label{fig:linear-harris}
\end{figure}

With both a quantified linear growth rate and a non-linear simulation, it is now possible to identify when the linear stage transitions to non-linear behavior. Like the Parker instability case, the Harris sheet has no equilibrium $B_x$-component, so we simply compare the simulation's $\max(B_x(t))$ to the linear prediction, i.e. $\max(B_x(t=0)) \exp(-\im\omega_\mathrm{max}t)$. This comparison is shown in Fig. \ref{fig:linear-harris}. As clarified by the inset, the simulation perfectly follows the linear theory in the very early stage, until around $t\sim 10$. After that time, the tearing instability transitions from exponential growth in the linear regime to linear growth in the Rutherford regime \citep{Rutherford1973} after $t\sim 200$, finally starting to saturate around $t\sim 300$ when further growth is counteracted by the perfectly conducting boundary conditions.

\subsubsection{Discrete Alfv\'en waves}
As a final example, we consider the following equilibrium representing a section of a coronal loop:
\begin{equation}
\begin{aligned}
    &\rho_0(r) = 1.0 - (1.0 - \delta)r^2,\\ &B_{\theta0}(r) = j_0  r  (r^4 - 3 r^2 + 3) / 6, \quad && B_{z0}(r) = 1,
\end{aligned}
\end{equation}
where $r\in[0.025,1]$. Pressure $p_0$ is then fixed by radial force balance up to a constant, for which we choose $p_0(1) = 5\times10^{-5}$. We include optically thin radiative losses, background heating, and parallel thermal conduction. This setup has been used to study the growth and damping of discrete Alfvén waves in coronal conditions by \cite{Keppens1993,Claes2020}. We use the cooling curve of \cite{Colgan2008}, modified at lower temperatures by \cite{Dalgarno1972} (\texttt{Colgan\_DM}). The background heating is set to exactly balance the initial losses. As is suitable for a fully ionized, hot corona, the perpendicular conduction $\kappa_\perp=0$, and the parallel conduction $\kappa_\parallel~=~8\times~10^{-7}~T^{5/2}$~ergs~cm$^{-1}$~s$^{-1}$~K$^{-1}$ is set to the Spitzer conductivity \citep{Spitzer2006}. We choose $j_0=0.5$, $\delta=0.2$, and consider modes with wavenumber $m=5$, $k=1$. We fix the unit density as $6 \times 10^{-15}$~g~$\text{cm}^{-3}$, unit magnetic field $25$~G, and unit length $10^{8}$~cm, representing typical conditions in coronal loops \citep{Reale2014}. In this case, working in dimensional variables is essential for the non-adiabatic terms. To simulate solar coronal conditions, we take a Helium abundance of 10\% in both \legolas{} and \amrvac{}.
\begin{figure}
    \centering
    \includegraphics[width=\linewidth]{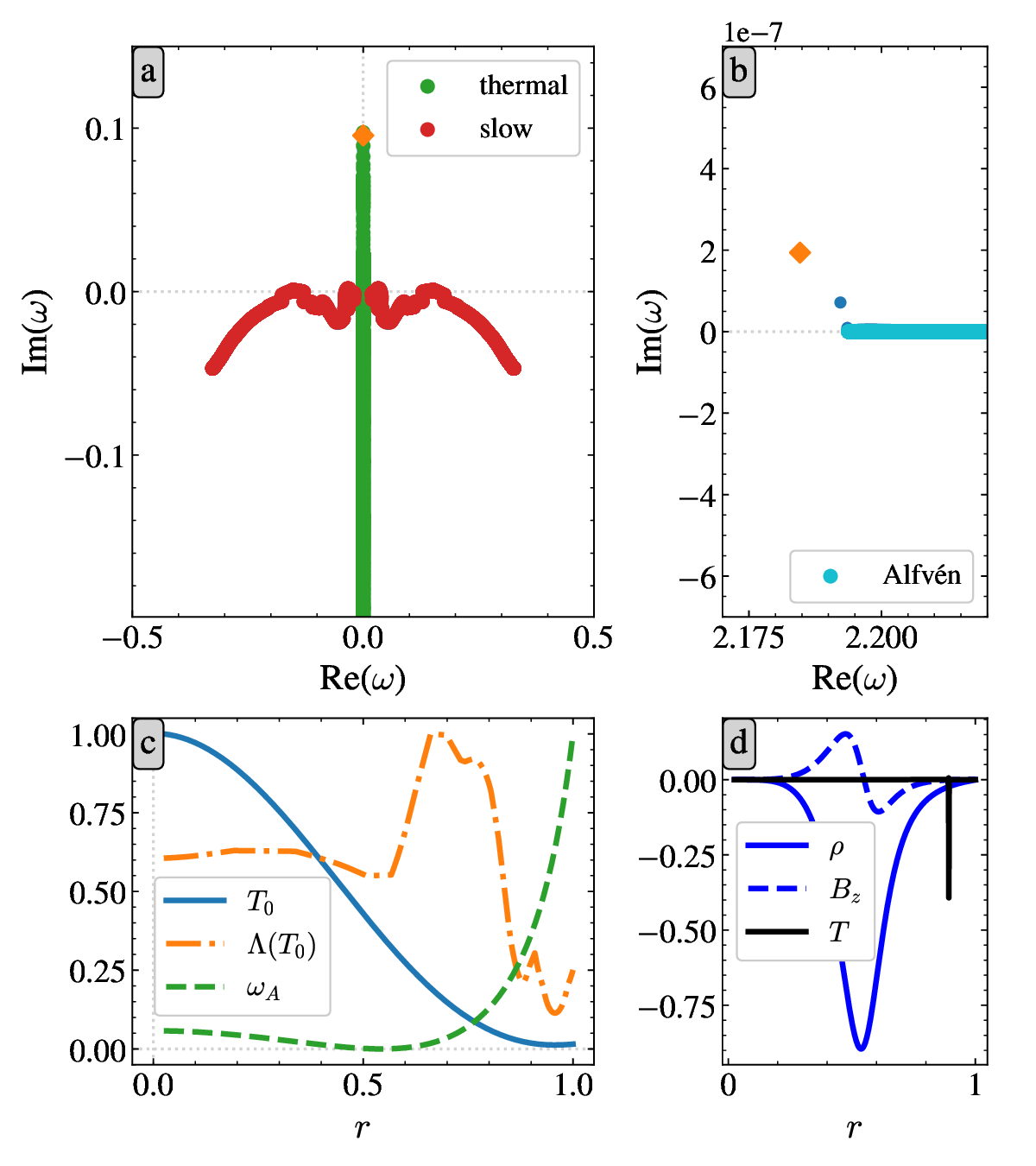}
    \caption{(a) Continuum thermal and slow eigenfrequencies. (b) Discrete and continuum Alfvén modes. The modes that we initialize are represented by orange diamonds. (c) Profiles of the Alfvén frequency, temperature, and cooling table evaluated at the equilibrium state. Rescaled and normalized for clarity. (d) Real part of the eigenfunctions of the most global discrete Alfvén mode for $m=5$ and $k=1.0$ (blue), and the most unstable thermal continuum mode (black).}
    \label{fig:legolas-discrete_alfven}
\end{figure}

Some equilibrium quantities are shown in Fig.~\ref{fig:legolas-discrete_alfven}, together with zoomed views of a spectrum obtained from a \legolas{} run with 500 grid points and the QR-Cholesky solution method. Panel~(a) shows how, because of the radiative losses, a thermal continuum appears, and the slow and fast mode frequencies obtain an imaginary component. A recent discussion of this thermal continuum can be found in \cite{TC2025}. Panel~(b) zooms in on the slightly unstable discrete Alfvén modes clustering towards the Alfvén continuum. We select the most global discrete mode at $\omega\approx2.185+1.940\times 10^{-7}\im$, and the most unstable thermal continuum mode at $\omega \approx 0.0955\im$, of which selected eigenfunctions are shown in Panel~(c). Note how the eigenfunctions peak around the radius where the radiative losses increase or decrease dramatically in Panel~(d). Because of the radial background variation, the Alfvén mode has mixed properties, and its temperature variation in this sensitive region is what causes it to extract energy from the background and become unstable. The continuum mode is analytically a singular solution, and the eigenmode here approximates a delta function.
\begin{figure}
    \centering
    \includegraphics[width=\linewidth]{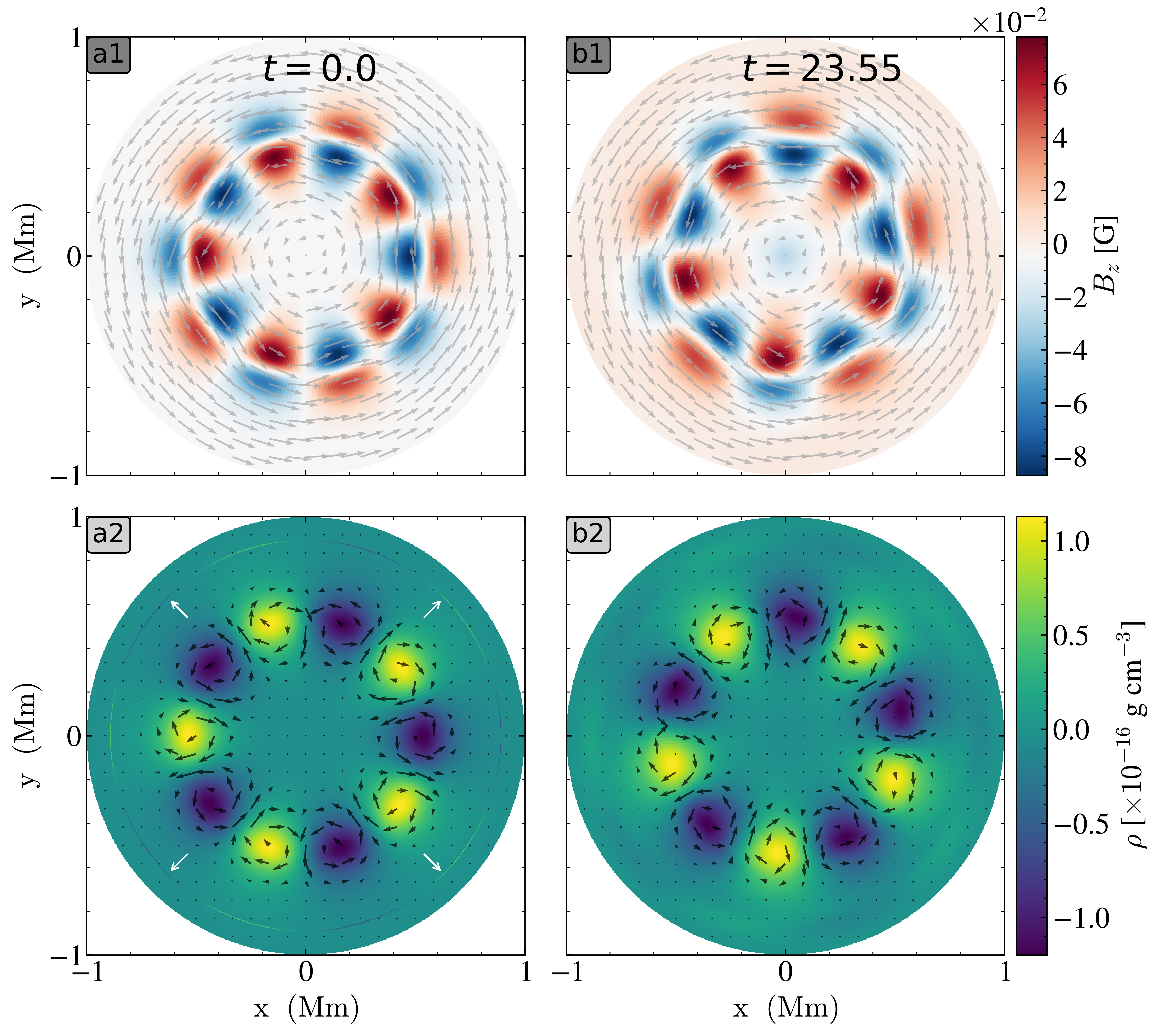}
    \caption{Perturbed $B_z$ and in-plane total magnetic field (first column) and perturbed $\rho$ and in-plane velocity field (second column) for $t = 0$ and $t=23$. Note how the very narrow thermal mode perturbation (indicated by white arrows) remains faintly present. Animated versions of this figure are available in the online manuscript.}
    \label{fig:amrvac-discrete_alfven}
\end{figure}

\begin{figure}
    \centering
    \includegraphics[width=\linewidth]{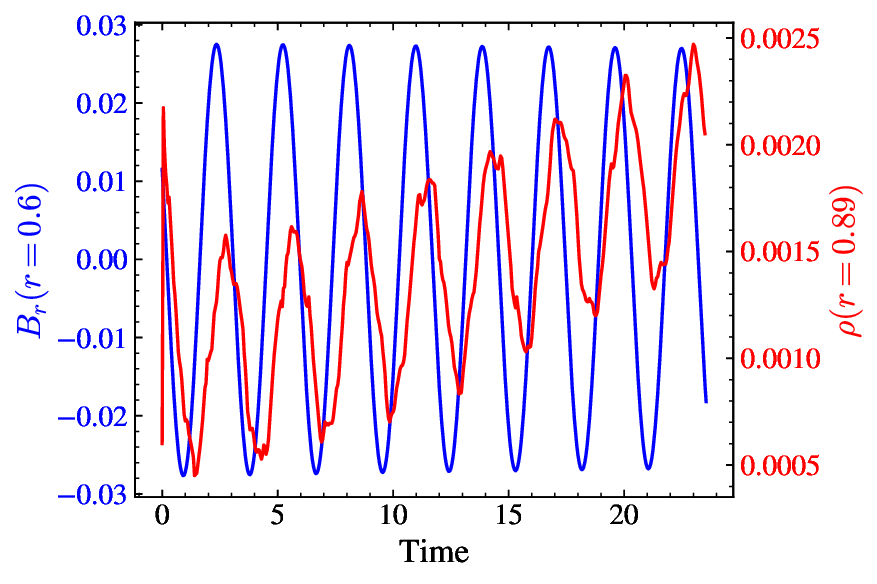}
    \caption{Evolution of the perturbed $B_z$ (blue) and perturbed $\rho$ (blue) at a fixed point $[0.89, 3.14, 3.0]$. The oscillation frequency $2.13$ at both locations is close to that of the initialized discrete Alfvén mode.}
    \label{fig:linear-discrete_alfven}
\end{figure}

This most global Alfvén mode and most unstable thermal mode is saved as initial condition for our \amrvac{} simulation using \gimli{} at a maximum value of $0.02\rho_0$ and with relative weights $0.7$ and $0.3$, respectively. We again use splitting for $\mathbf{B_0}$, $\mathbf{j_0}$, $\rho_0$, and $p_0$, and apply the same boundary conditions as in Sec.~\ref{sec:tokamak}. We use a resolution of $256\times256\times128$ on a domain $r\in[0,1]$, $\theta\in[0,2\pi]$, and $z\in[0,\lambda]$, where $\lambda=2\pi$. The initial condition is again advanced using a threestep time integrator and HLL flux scheme with the cada3 limiter. For the optically thin radiative losses, \amrvac{} uses an exact integration scheme, and we use hyperbolic thermal conduction with saturation \citep{Zhou2025}. Figure~\ref{fig:amrvac-discrete_alfven} shows the initial condition for the perturbations in $B_z$ and $\rho$ as well as a later snapshot. The Alfvén mode propagates relatively unchanged in time. Initially, the thermal mode is shown as a very sharp perturbation. It quickly spreads out into a weak mode that is damped and converted into an oscillating mode with small amplitude. At $r=0.89$, where the thermal mode peaks, $\theta=\pi/4$, an analysis of the vertical mode structure at $t=23.55$ reveals a dominant wavenumber $k=1.0$ as expected, but with increased power in $k=2$ as well. In Fig.~\ref{fig:linear-discrete_alfven}, we show the temporal evolution of $B_z$ (blue) at a fixed point $(0.6, 3.14, 3)$ around where the eigenfunctions peak. An oscillating mode is clearly excited by the initial condition. A Fourier analysis reveals a dominant frequency of $\omega=2.13$, which is close to the analytical frequency of $\omega\approx2.185$. The evolution of the perturbed $\rho$ at $r=0.89$, where the thermal mode is initialized, is shown in red. There, an oscillating mode is excited with the same frequency $2.13$ that grows more slowly than exponentially. This growth is only present in the thermodynamic variables and not the magnetic field, which is a signature of the initialized thermal mode. We checked that the growth remains roughly linear by extending the simulation to $t=80$. Follow-up research can meaningfully look into the far non-linear consequences of having both unstable thermal continuum modes, as well as discrete Alfv\'en modes, in similar setups as used for the demonstration purpose here.

\section{Outlook}\label{sec:outlook}
In this work, we presented \gimli{}, a Python toolkit linking the linear MHD spectroscopy code \legolas{} and the non-linear solver \amrvac{}. Using \gimli{}'s functionalities, any analytical prescription of a one-dimensionally varying MHD state can be defined to generate \legolas{} and \amrvac{} user modules, as well as their complementary parameter files. After running the \legolas{} code, linear solutions can be selected and added as initial perturbations to \amrvac{} simulations. This process was demonstrated for four unstable equilibria, two in ideal MHD and two including non-ideal effects, in both Cartesian and cylindrical geometries.

Uniting linear and non-linear codes, \gimli{} provides a framework to perform studies in both regimes to obtain a complete picture of a system's behavior. Notably, it facilitates a direct comparison between the linearly predicted growth rate and the non-linear evolution, allowing for the determination of the onset time of non-linear effects. Furthermore, as shown by \citet{DeJonghe2026a}, the method of initializing non-linear simulations with linear solutions provides a tangible computational advantage. However, as demonstrated in that same work, omitting certain modes can significantly affect the non-linear evolution. This makes \gimli{} a valuable tool to understand the role of different linear modes in observed non-linear properties, and how various modes interact non-linearly. Eventually, it would be possible to complement these simulations with linear mode projection of fully turbulent states onto the \legolas{} eigenmode bases to track the time evolution of the most important modes \citep{Pueschel2016, Fraser2018}. A combined linear-nonlinear approach has promising applications to accretion disks, where recent results on non-axisymmetric modes invite such an analysis \citep{Goedbloed2022, Brughmans2024, Brughmans2025}. Extremely high resolution simulations of isolated MHD modes using GPU codes like \agile{}, complemented with the linear analysis of evolving states as in \citet{DeJonghe2026b}, will help to disentangle the role of different ideal MHD instabilities in dynamo amplification and the non-linear evolution to turbulence.

\section*{Data availability}
The results presented in this work were obtained with \legolas{} v2.3.1 \citep[\url{https://legolas.science}]{Claes2020, DeJonghe2022, Claes2023} and \amrvac{} v3.3 \citep[\url{https://amrvac.org}]{Porth2014, Xia2018, Keppens2023}. Visualizations were performed with \yt{} v4.4.0 \citep[\url{https://yt-project.org}]{Turk2011} and \pyvista{} v0.48.4\citep[\url{https://pyvista.org/}]{Sullivan2019}.

\begin{acknowledgements}
JDJ acknowledges funding by the UK's Science and Technology Facilities Council (STFC) Consolidated Grant ST/W001195/1 and the Research Foundation - Flanders (FWO) fellowship 1225625N. 
 RK acknowledges funding from the KU Leuven C1 project C16/24/010 UnderRadioSun and the Research Foundation Flanders FWO project G0B9923N Helioskill. Computational resources and services used for the 3D simulations were provided by the VSC (Flemish Supercomputer Center), funded by the Research Foundation Flanders (FWO) and the Flemish Government - department EWI. This research has made use of NASA's Astrophysics Data System Bibliographic Services.
\end{acknowledgements}

\bibliography{bibliography.bib}{}

@ARTICLE{TC2025,
       author = {{Keppens}, Rony and {De Jonghe}, Jordi and {Kelly}, Adrian and {Brughmans}, Nicolas and {Goedbloed}, Hans},
        title = "{The Hydrodynamic Thermal Continuum, with Applications to Stratified Atmospheres and 1D Coronal Loop Models}",
      journal = {\apj},
         year = 2025,
        month = aug,
       volume = {989},
       number = {1},
          eid = {51},
        pages = {51},
          doi = {10.3847/1538-4357/adea43},
archivePrefix = {arXiv},
       eprint = {2506.23591},
 primaryClass = {astro-ph.SR},
       adsurl = {https://ui.adsabs.harvard.edu/abs/2025ApJ...989...51K}
}

@ARTICLE{fieldtheory2016,
       author = {{Keppens}, R. and {Demaerel}, T.},
        title = "{Stability of ideal MHD configurations. I. Realizing the generality of the G operator}",
      journal = {Physics of Plasmas},
         year = 2016,
        month = dec,
       volume = {23},
       number = {12},
          eid = {122117},
        pages = {122117},
          doi = {10.1063/1.4971811},
       adsurl = {https://ui.adsabs.harvard.edu/abs/2016PhPl...23l2117K}
}

@ARTICLE{phoenix2007,
       author = {{Blokland}, J.~W.~S. and {van der Holst}, B. and {Keppens}, R. and {Goedbloed}, J.~P.},
        title = "{PHOENIX: MHD spectral code for rotating laboratory and gravitating astrophysical plasmas}",
      journal = {Journal of Computational Physics},
         year = 2007,
        month = sep,
       volume = {226},
        pages = {509-533},
          doi = {10.1016/j.jcp.2007.04.018},
       adsurl = {https://ui.adsabs.harvard.edu/abs/2007JCoPh.226..509B}
}

@article{Kerner1985,
    author = {{Kerner}, W. and {Lerbinger}, K. and {Gruber}, R. and {Tsunematsu}, T.},
    title = "{Normal mode analysis for resistive cylindrical plasmas}",
    journal = {Computer Physics Communications},
    year = 1985,
    month = may,
    volume = {36},
    number = {3},
    pages = {225-240},
    doi = {10.1016/0010-4655(85)90053-0},
    adsurl = {https://ui.adsabs.harvard.edu/abs/1985CoPhC..36..225K}
}

@article{Nijboer1997,
    author = {{Nijboer}, R.~J. and {Holst}, B. v. d. and {Poedts}, S. and {Goedbloed}, J.~P.},
    title = "{Calculating magnetohydrodynamic flow spectra}",
    journal = {Computer Physics Communications},
    year = 1997,
    month = oct,
    volume = {106},
    number = {1},
    pages = {39-52},
    doi = {10.1016/S0010-4655(97)00082-9},
    adsurl = {https://ui.adsabs.harvard.edu/abs/1997CoPhC.106...39N}
}

@article{Arber2001,
    author = {{Arber}, T.~D. and {Longbottom}, A.~W. and {Gerrard}, C.~L. and {Milne}, A.~M.},
    title = "{A Staggered Grid, Lagrangian-Eulerian Remap Code for 3-D MHD Simulations}",
    journal = {Journal of Computational Physics},
    year = 2001,
    month = jul,
    volume = {171},
    number = {1},
    pages = {151-181},
    doi = {10.1006/jcph.2001.6780},
    adsurl = {https://ui.adsabs.harvard.edu/abs/2001JCoPh.171..151A}
}

@article{Mignone2007,
    author = {{Mignone}, A. and {Bodo}, G. and {Massaglia}, S. and {Matsakos}, T. and {Tesileanu}, O. and {Zanni}, C. and {Ferrari}, A.},
    title = "{PLUTO: A Numerical Code for Computational Astrophysics}",
    journal = {\apjs},
    year = 2007,
    month = may,
    volume = {170},
    number = {1},
    pages = {228-242},
    doi = {10.1086/513316},
    archivePrefix = {arXiv},
    eprint = {astro-ph/0701854},
    primaryClass = {astro-ph},
    adsurl = {https://ui.adsabs.harvard.edu/abs/2007ApJS..170..228M}
}

@article{Turk2011,
    author = {{Turk}, Matthew J. and {Smith}, Britton D. and {Oishi}, Jeffrey S. and {Skory}, Stephen and {Skillman}, Samuel W. and {Abel}, Tom and {Norman}, Michael L.},
    title = "{yt: A Multi-code Analysis Toolkit for Astrophysical Simulation Data}",
    journal = {\apjs},
    year = 2011,
    month = jan,
    volume = {192},
    number = {1},
    eid = {9},
    pages = {9},
    doi = {10.1088/0067-0049/192/1/9},
    archivePrefix = {arXiv},
    eprint = {1011.3514},
    primaryClass = {astro-ph.IM},
    adsurl = {https://ui.adsabs.harvard.edu/abs/2011ApJS..192....9T}
}

@article{Mignone2012,
    author = {{Mignone}, A. and {Zanni}, C. and {Tzeferacos}, P. and {van Straalen}, B. and {Colella}, P. and {Bodo}, G.},
    title = "{The PLUTO Code for Adaptive Mesh Computations in Astrophysical Fluid Dynamics}",
    journal = {\apjs},
    year = 2012,
    month = jan,
    volume = {198},
    number = {1},
    eid = {7},
    pages = {7},
    doi = {10.1088/0067-0049/198/1/7},
    archivePrefix = {arXiv},
    eprint = {1110.0740},
    primaryClass = {astro-ph.HE},
    adsurl = {https://ui.adsabs.harvard.edu/abs/2012ApJS..198....7M}
}

@article{Porth2014,
    author = {{Porth}, O. and {Xia}, C. and {Hendrix}, T. and {Moschou}, S.~P. and {Keppens}, R.},
    title = "{MPI-AMRVAC for Solar and Astrophysics}",
    journal = {\apjs},
    year = 2014,
    month = sep,
    volume = {214},
    number = {1},
    eid = {4},
    pages = {4},
    doi = {10.1088/0067-0049/214/1/4},
    archivePrefix = {arXiv},
    eprint = {1407.2052},
    primaryClass = {astro-ph.IM},
    adsurl = {https://ui.adsabs.harvard.edu/abs/2014ApJS..214....4P}
}

@article{Meurer2017,
    title = {SymPy: symbolic computing in Python},
    author = {Meurer, Aaron and Smith, Christopher P. and Paprocki, Mateusz and \v{C}ert\'{i}k, Ond\v{r}ej and Kirpichev, Sergey B. and Rocklin, Matthew and Kumar, AMiT and Ivanov, Sergiu and Moore, Jason K. and Singh, Sartaj and Rathnayake, Thilina and Vig, Sean and Granger, Brian E. and Muller, Richard P. and Bonazzi, Francesco and Gupta, Harsh and Vats, Shivam and Johansson, Fredrik and Pedregosa, Fabian and Curry, Matthew J. and Terrel, Andy R. and Rou\v{c}ka, \v{S}t\v{e}p\'{a}n and Saboo, Ashutosh and Fernando, Isuru and Kulal, Sumith and Cimrman, Robert and Scopatz, Anthony},
    year = 2017,
    month = jan,
    volume = 3,
    pages = {e103},
    journal = {PeerJ Computer Science},
    issn = {2376-5992},
    url = {https://doi.org/10.7717/peerj-cs.103},
    doi = {10.7717/peerj-cs.103}
}

@article{Xia2018,
    author = {{Xia}, C. and {Teunissen}, J. and {El Mellah}, I. and {Chan{\'e}}, E. and {Keppens}, R.},
    title = "{MPI-AMRVAC 2.0 for Solar and Astrophysical Applications}",
    journal = {\apjs},
    year = 2018,
    month = feb,
    volume = {234},
    number = {2},
    eid = {30},
    pages = {30},
    doi = {10.3847/1538-4365/aaa6c8},
    archivePrefix = {arXiv},
    eprint = {1710.06140},
    primaryClass = {astro-ph.SR},
    adsurl = {https://ui.adsabs.harvard.edu/abs/2018ApJS..234...30X}
}

@article{Goedbloed2018a,
    author = {{Goedbloed}, J.~P.},
    title = "{The Spectral Web of stationary plasma equilibria. I. General theory}",
    journal = {Physics of Plasmas},
    year = 2018,
    month = mar,
    volume = {25},
    number = {3},
    eid = {032109},
    pages = {032109},
    doi = {10.1063/1.5019831},
    adsurl = {https://ui.adsabs.harvard.edu/abs/2018PhPl...25c2109G}
}

@article{Goedbloed2018b,
    author = {{Goedbloed}, J.~P.},
    title = "{The Spectral Web of stationary plasma equilibria. II. Internal modes}",
    journal = {Physics of Plasmas},
    year = 2018,
    month = mar,
    volume = {25},
    number = {3},
    eid = {032110},
    pages = {032110},
    doi = {10.1063/1.5019838},
    adsurl = {https://ui.adsabs.harvard.edu/abs/2018PhPl...25c2110G}
}

@book{Goedbloed2019,
    author = {{Goedbloed}, Hans and {Keppens}, Rony and {Poedts}, Stefaan},
    title = "{Magnetohydrodynamics of Laboratory and Astrophysical Plasmas}",
    year = 2019,
    doi = {10.1017/9781316403679},
    adsurl = {https://ui.adsabs.harvard.edu/abs/2019mlap.book.....G}
}

@article{Stone2020,
    author = {{Stone}, James M. and {Tomida}, Kengo and {White}, Christopher J. and {Felker}, Kyle G.},
    title = "{The Athena++ Adaptive Mesh Refinement Framework: Design and Magnetohydrodynamic Solvers}",
    journal = {\apjs},
    year = 2020,
    month = jul,
    volume = {249},
    number = {1},
    eid = {4},
    pages = {4},
    doi = {10.3847/1538-4365/ab929b},
    archivePrefix = {arXiv},
    eprint = {2005.06651},
    primaryClass = {astro-ph.IM},
    adsurl = {https://ui.adsabs.harvard.edu/abs/2020ApJS..249....4S}
}

@article{Claes2020,
    author = {{Claes}, Niels and {De Jonghe}, Jordi and {Keppens}, Rony},
    title = "{Legolas: A Modern Tool for Magnetohydrodynamic Spectroscopy}",
    journal = {\apjs},
    year = 2020,
    month = dec,
    volume = {251},
    number = {2},
    eid = {25},
    pages = {25},
    doi = {10.3847/1538-4365/abc5c4},
    archivePrefix = {arXiv},
    eprint = {2010.14148},
    primaryClass = {astro-ph.SR},
    adsurl = {https://ui.adsabs.harvard.edu/abs/2020ApJS..251...25C}
}

@article{Claes2021,
    author = {{Claes}, Niels and {Keppens}, Rony},
    title = "{Magnetohydrodynamic Spectroscopy of a Non-adiabatic Solar Atmosphere}",
    journal = {\solphys},
    year = 2021,
    month = sep,
    volume = {296},
    number = {9},
    eid = {143},
    pages = {143},
    doi = {10.1007/s11207-021-01894-2},
    archivePrefix = {arXiv},
    eprint = {2108.09467},
    primaryClass = {astro-ph.SR},
    adsurl = {https://ui.adsabs.harvard.edu/abs/2021SoPh..296..143C}
}

@article{DeJonghe2022,
    author = {{De Jonghe}, J. and {Claes}, N. and {Keppens}, R.},
    title = "{Legolas: magnetohydrodynamic spectroscopy with viscosity and Hall current}",
    journal = {Journal of Plasma Physics},
    year = 2022,
    month = jun,
    volume = {88},
    number = {3},
    eid = {905880321},
    pages = {905880321},
    doi = {10.1017/S0022377822000617},
    archivePrefix = {arXiv},
    eprint = {2206.07377},
    primaryClass = {physics.plasm-ph},
    adsurl = {https://ui.adsabs.harvard.edu/abs/2022JPlPh..88c9021D}
}

@article{Claes2023,
    author = {{Claes}, Niels and {Keppens}, Rony},
    title = "{Legolas 2.0: Improvements and extensions to an MHD spectroscopic framework}",
    journal = {Computer Physics Communications},
    year = 2023,
    month = oct,
    volume = {291},
    eid = {108856},
    pages = {108856},
    doi = {10.1016/j.cpc.2023.108856},
    archivePrefix = {arXiv},
    eprint = {2307.10145},
    primaryClass = {astro-ph.IM},
    adsurl = {https://ui.adsabs.harvard.edu/abs/2023CoPhC.29108856C}
}

@article{Keppens2023,
    author = {{Keppens}, R. and {Popescu Braileanu}, B. and {Zhou}, Y. and {Ruan}, W. and {Xia}, C. and {Guo}, Y. and {Claes}, N. and {Bacchini}, F.},
    title = "{MPI-AMRVAC 3.0: Updates to an open-source simulation framework}",
    journal = {\aap},
    year = 2023,
    month = may,
    volume = {673},
    eid = {A66},
    pages = {A66},
    doi = {10.1051/0004-6361/202245359},
    archivePrefix = {arXiv},
    eprint = {2303.03026},
    primaryClass = {astro-ph.IM},
    adsurl = {https://ui.adsabs.harvard.edu/abs/2023A&A...673A..66K}
}

@article{DeJonghe2024,
    author = {{De Jonghe}, J. and {Keppens}, R.},
    title = "{Modification of the resistive tearing instability with Joule heating by shear flow}",
    journal = {Physics of Plasmas},
    year = 2024,
    month = mar,
    volume = {31},
    number = {3},
    eid = {032106},
    pages = {032106},
    doi = {10.1063/5.0180535},
    archivePrefix = {arXiv},
    eprint = {2402.12005},
    primaryClass = {physics.plasm-ph},
    adsurl = {https://ui.adsabs.harvard.edu/abs/2024PhPl...31c2106D}
}

@article{Brughmans2024,
    author = {{Brughmans}, Nicolas and {Keppens}, Rony and {Goedbloed}, Hans},
    title = "{Parametric Survey of Nonaxisymmetric Accretion Disk Instabilities: Magnetorotational Instability to Super-Alfv{\'e}nic Rotational Instability}",
    journal = {\apj},
    year = 2024,
    month = jun,
    volume = {968},
    number = {1},
    eid = {19},
    pages = {19},
    doi = {10.3847/1538-4357/ad3d52},
    archivePrefix = {arXiv},
    eprint = {2404.06925},
    primaryClass = {astro-ph.HE},
    adsurl = {https://ui.adsabs.harvard.edu/abs/2024ApJ...968...19B}
}

@article{DeJonghe2025,
    author = {{De Jonghe}, Jordi and {Sen}, Samrat},
    title = "{The coupled tearing-thermal instability in coronal current sheets from the linear to the non-linear stage}",
    journal = {\mnras},
    year = 2025,
    month = feb,
    volume = {536},
    number = {4},
    pages = {3308-3321},
    doi = {10.1093/mnras/stae2740},
    archivePrefix = {arXiv},
    eprint = {2412.07427},
    primaryClass = {astro-ph.SR},
    adsurl = {https://ui.adsabs.harvard.edu/abs/2025MNRAS.536.3308D}
}

@article{DeJonghe2026a,
    author = {{De Jonghe}, J. and {Russell}, A.~J.~B.},
    title = "{Eigenmode initialisation of 2D (magneto)hydrodynamic simulations}",
    journal = {\aap},
    year = 2026,
    month = apr,
    volume = {708},
    eid = {A138},
    pages = {A138},
    doi = {10.1051/0004-6361/202452721},
    archivePrefix = {arXiv},
    eprint = {2602.23849},
    primaryClass = {astro-ph.IM},
    adsurl = {https://ui.adsabs.harvard.edu/abs/2026A&A...708A.138D}
}

@misc{DeJonghe2026b,
      title={Streamer slab eigenmode analysis with the Legolas code}, 
      author={Jordi {De Jonghe} and Daria Sorokina and Tom {Van Doorsselaere}},
      year={2026},
      eprint={2609.24442},
      archivePrefix={arXiv},
      primaryClass={astro-ph.SR},
      url={https://arxiv.org/abs/2609.24442}
}

@article{Kelly2026,
    author = {{Kelly}, Adrian and {Keppens}, Rony and {De Jonghe}, Jordi},
    title = "{Thermal instability in coronal loops: Linking eigenvalue spectra to time-dependent evolution}",
    journal = {\aap},
    year = 2026,
    month = may,
    volume = {710},
    eid = {A19},
    pages = {A19},
    doi = {10.1051/0004-6361/202659934},
    archivePrefix = {arXiv},
    eprint = {2604.24315},
    primaryClass = {astro-ph.SR},
    adsurl = {https://ui.adsabs.harvard.edu/abs/2026A&A...710A..19K}
}

@ARTICLE{Keppens1993,
       author = {{Keppens}, Rony and {van der Linden}, Ronald A.~M. and {Goossens}, Marcel},
        title = "{Non-adiabatic discrete Alfv{\'e}n waves in coronal loops and prominences}",
      journal = {\solphys},
         year = 1993,
        month = apr,
       volume = {144},
       number = {2},
        pages = {267-281},
          doi = {10.1007/BF00627593},
       adsurl = {https://ui.adsabs.harvard.edu/abs/1993SoPh..144..267K}
}

@article{Colgan2008,
issn = {0004-637X},
journal = {\apj},
pages = {585--592},
volume = {689},
publisher = {IOP Publishing},
number = {1},
year = {2008},
title = "{Radiative Losses of Solar Coronal Plasmas}",
author = {Colgan, J and Abdallah, Jr., J and Sherrill, M. E and Foster, M and Fontes, C. J and Feldman, U},
OPTcomm={Maybe read this paper as background on cooling curve.}
}

@ARTICLE{Dalgarno1972,
       author = {{Dalgarno}, A. and {McCray}, R.~A.},
        title = "{Heating and Ionization of HI Regions}",
      journal = {Annual Review of Astronomy and Astrophysics},
         year = 1972,
        month = jan,
       volume = {10},
        pages = {375},
          doi = {10.1146/annurev.aa.10.090172.002111},
       adsurl = {https://ui.adsabs.harvard.edu/abs/1972ARA&A..10..375D}
}

@ARTICLE{Brughmans2025,
       author = {{Brughmans}, Nicolas and {Keppens}, Rony},
        title = "{A visual approach to global accretion disc instabilities}",
      journal = {\mnras},
         year = 2025,
        month = sep,
       volume = {542},
       number = {2},
        pages = {1347-1363},
          doi = {10.1093/mnras/staf1265},
archivePrefix = {arXiv},
       eprint = {2507.22672},
 primaryClass = {astro-ph.GA},
       adsurl = {https://ui.adsabs.harvard.edu/abs/2025MNRAS.542.1347B}
}

@book{Spitzer2006,
  title={Physics of fully ionized gases},
  author={Spitzer, Lyman},
  year={2006},
  publisher={Courier Corporation}
}

@ARTICLE{Hatch2016,
       author = {{Hatch}, D.~R. and {Jenko}, F. and {Ba{\~n}{\'o}n Navarro}, A. and {Bratanov}, V. and {Terry}, P.~W. and {Pueschel}, M.~J.},
        title = "{Linear signatures in nonlinear gyrokinetics: interpreting turbulence with pseudospectra}",
      journal = {New Journal of Physics},
         year = 2016,
        month = jul,
       volume = {18},
       number = {7},
          eid = {075018},
        pages = {075018},
          doi = {10.1088/1367-2630/18/7/075018},
       adsurl = {https://ui.adsabs.harvard.edu/abs/2016NJPh...18g5018H}
}

@ARTICLE{Pueschel2016,
       author = {{Pueschel}, M.~J. and {Faber}, B.~J. and {Citrin}, J. and {Hegna}, C.~C. and {Terry}, P.~W. and {Hatch}, D.~R.},
        title = "{Stellarator Turbulence: Subdominant Eigenmodes and Quasilinear Modeling}",
      journal = {\prl},
         year = 2016,
        month = feb,
       volume = {116},
       number = {8},
          eid = {085001},
        pages = {085001},
          doi = {10.1103/PhysRevLett.116.085001},
       adsurl = {https://ui.adsabs.harvard.edu/abs/2016PhRvL.116h5001P}
}

@ARTICLE{Fraser2018,
       author = {{Fraser}, A.~E. and {Pueschel}, M.~J. and {Terry}, P.~W. and {Zweibel}, E.~G.},
        title = "{Role of stable modes in driven shear-flow turbulence}",
      journal = {Physics of Plasmas},
         year = 2018,
        month = dec,
       volume = {25},
       number = {12},
          eid = {122303},
        pages = {122303},
          doi = {10.1063/1.5049580},
archivePrefix = {arXiv},
       eprint = {1807.09280},
 primaryClass = {physics.plasm-ph},
       adsurl = {https://ui.adsabs.harvard.edu/abs/2018PhPl...25l2303F}
}

@ARTICLE{Porth2026,
       author = {{Porth}, Oliver and {Kelly}, Adrian and {Willocx}, Olaf and {Wu}, Hao and {Vos}, Jesse and {Zhou}, Yuhao and {Olivares S{\'a}nchez}, H{\'e}ctor R. and {Oostrum}, Leon and {Hidding}, Johan and {Azizi}, Victor and {Xia}, Chun and {Keppens}, Rony and {Teunissen}, Jannis},
        title = "{Astrophysics on GPUs: introducing AGILE 1.0}",
      journal = {arXiv e-prints},
         year = 2026,
        month = jul,
          eid = {arXiv:2607.19277},
        pages = {arXiv:2607.19277},
          doi = {10.48550/arXiv.2607.19277},
archivePrefix = {arXiv},
       eprint = {2607.19277},
 primaryClass = {astro-ph.IM},
       adsurl = {https://ui.adsabs.harvard.edu/abs/2026arXiv260719277P}
}

@ARTICLE{Zhou2025,
       author = {{Zhou}, Yuhao and {Li}, Xiaohong and {Jenkins}, Jack M. and {Hong}, Jie and {Keppens}, Rony},
        title = "{Frozen-field Modeling of Coronal Condensations with MPI-AMRVAC. II. Optimization and Application in 3D Models}",
      journal = {\apj},
         year = 2025,
        month = jan,
       volume = {978},
       number = {1},
          eid = {72},
        pages = {72},
          doi = {10.3847/1538-4357/ad96af},
archivePrefix = {arXiv},
       eprint = {2411.16415},
 primaryClass = {astro-ph.SR},
       adsurl = {https://ui.adsabs.harvard.edu/abs/2025ApJ...978...72Z}
}

@ARTICLE{Reale2014,
       author = {{Reale}, Fabio},
        title = "{Coronal Loops: Observations and Modeling of Confined Plasma}",
      journal = {Living Reviews in Solar Physics},
         year = 2014,
        month = dec,
       volume = {11},
       number = {1},
          eid = {4},
        pages = {4},
          doi = {10.12942/lrsp-2014-4},
       adsurl = {https://ui.adsabs.harvard.edu/abs/2014LRSP...11....4R}
}

@article{Sullivan2019,
  title={PyVista: 3D plotting and mesh analysis through a streamlined interface for the Visualization Toolkit (VTK)},
  author={Sullivan, Cory and Kaszynski, Alexander},
  journal={Journal of Open Source Software},
  volume={4},
  number={37},
  pages={1450},
  year={2019},
  publisher={The Open Journal}
}

@article{Goedbloed2022,
  title={The super-Alfv{\'e}nic rotational instability in accretion disks about black holes},
  author={Goedbloed, Hans and Keppens, Rony},
  journal={\apjs},
  volume={259},
  number={2},
  pages={65},
  year={2022},
  publisher={The American Astronomical Society}
}

@ARTICLE{Rutherford1973,
       author = {{Rutherford}, P.~H.},
        title = "{Nonlinear growth of the tearing mode}",
      journal = {Physics of Fluids},
         year = 1973,
        month = nov,
       volume = {16},
       number = {11},
        pages = {1903-1908},
          doi = {10.1063/1.1694232},
       adsurl = {https://ui.adsabs.harvard.edu/abs/1973PhFl...16.1903R}
}
\bibliographystyle{aa}

\end{document}